\pdfoutput=1
\documentclass[12pt,a4paper]{article}

\usepackage{ifthen} 
\usepackage{rotating}
\usepackage{dashrule}
\usepackage{array} 
\usepackage{dcolumn}
\usepackage{comment}
\usepackage{xcolor}

\usepackage{graphpap} 
\usepackage{rotating} 
\usepackage{tikz}
\usepackage{xcolor}
\usepackage{longtable} 

\usepackage{lscape}
\usepackage{multirow} 
\usepackage{mathrsfs}
\usepackage{mathtools} 
\usetikzlibrary{patterns}
\usepackage{nicefrac} 
\usepackage{transparent}
\usepackage{afterpage} 
\usepackage{comment} 
\usepackage[normalem]{ulem}
\usepackage{cancel} 

\newboolean{pdflatex}
\setboolean{pdflatex}{true} 

\newboolean{articletitles}
\setboolean{articletitles}{true} 

\newboolean{uprightparticles}
\setboolean{uprightparticles}{True} 

\def\paperauthors{LHCb collaboration} 
\def\paperasciititle{Observation of Bs->X(3872)pipi decay} 
\def\papertitle{Observation of 
the~\mbox{$\decay{\Bs}{\chicone(3872)\pip\pim}$}~decay} 
\def\paperkeywords{{High Energy Physics}, {LHCb}} 
\def\papercopyright{\the\year\ CERN for the benefit of the LHCb collaboration} 
\def\paperlicence{CC BY 4.0 licence}
\def\paperlicenceurl{https://creativecommons.org/licenses/by/4.0/}

\DeclareMathOperator*{\bigplus}{\scalerel*{+}{\sum}}
\usepackage{scalerel}

\makeatletter
\g@addto@macro\bfseries{\boldmath}
\makeatother

\usepackage[top=1in, bottom=1.25in, left=1in, right=1in]{geometry}

\usepackage{microtype}
\usepackage{lineno}  
\usepackage{xspace} 
\usepackage{caption} 

\usepackage{graphicx}  
\usepackage{color}
\usepackage{colortbl}
\graphicspath{{./figs/}} 

\usepackage{amsmath} 
\usepackage{amssymb}
\usepackage{amsfonts}
\usepackage{upgreek} 

\newcommand*\patchAmsMathEnvironmentForLineno[1]{%
\expandafter\let\csname old#1\expandafter\endcsname\csname #1\endcsname
\expandafter\let\csname oldend#1\expandafter\endcsname\csname
end#1\endcsname
 \renewenvironment{#1}%
   {\linenomath\csname old#1\endcsname}%
   {\csname oldend#1\endcsname\endlinenomath}%
}
\newcommand*\patchBothAmsMathEnvironmentsForLineno[1]{%
  \patchAmsMathEnvironmentForLineno{#1}%
  \patchAmsMathEnvironmentForLineno{#1*}%
}
\AtBeginDocument{%
\patchBothAmsMathEnvironmentsForLineno{equation}%
\patchBothAmsMathEnvironmentsForLineno{align}%
\patchBothAmsMathEnvironmentsForLineno{flalign}%
\patchBothAmsMathEnvironmentsForLineno{alignat}%
\patchBothAmsMathEnvironmentsForLineno{gather}%
\patchBothAmsMathEnvironmentsForLineno{multline}%
\patchBothAmsMathEnvironmentsForLineno{eqnarray}%
}

\usepackage{hyperxmp}

\usepackage[pdftex,
            pdfauthor={\paperauthors},
            pdftitle={\paperasciititle},
            pdfkeywords={\paperkeywords},
            pdfcopyright={Copyright (C) \papercopyright},
            pdflicenseurl={\paperlicenceurl}]{hyperref}

\usepackage[colorinlistoftodos,textsize=scriptsize]{todonotes}

\usepackage[bottom,flushmargin,hang,multiple]{footmisc}

\usepackage[all]{hypcap} 

\usepackage{xspace} 
\usepackage{upgreek}

\def\lhcb   {\mbox{LHCb}\xspace}

\def\MagUp {\mbox{\em Mag\kern -0.05em Up}\xspace}

\ifthenelse{\boolean{uprightparticles}}%
{

 \def\Pmu         {\ensuremath{\upmu}\xspace}

 \def\Ppi         {\ensuremath{\uppi}\xspace}                 
                  
 \def\Prho        {\ensuremath{\uprho}\xspace}

 \def\Pphi        {\ensuremath{\upphi}\xspace}                 
                  
 \def\Pchi        {\ensuremath{\upchi}\xspace}                 
 \def\Ppsi        {\ensuremath{\uppsi}\xspace}

 \def\PDelta      {\ensuremath{\Delta}\xspace}                 
 \def\PXi         {\ensuremath{\Xi}\xspace}                 
 \def\PLambda     {\ensuremath{\Lambda}\xspace}                 
 \def\PSigma      {\ensuremath{\Sigma}\xspace}                 
 \def\POmega      {\ensuremath{\Omega}\xspace}                 
 \def\PUpsilon    {\ensuremath{\Upsilon}\xspace}
 \let\oldPi\Pi
 \def\PPi         {\ensuremath{\oldPi}\xspace}

 \def\PB      {\ensuremath{\mathrm{B}}\xspace}                 
                  
 \def\PD      {\ensuremath{\mathrm{D}}\xspace}

 \def\PJ      {\ensuremath{\mathrm{J}}\xspace}                 
 \def\PK      {\ensuremath{\mathrm{K}}\xspace}

 \def\PS      {\ensuremath{\mathrm{S}}\xspace}

 \def\Pb      {\ensuremath{\mathrm{b}}\xspace}                 
 \def\Pc      {\ensuremath{\mathrm{c}}\xspace}

 \def\Pf      {\ensuremath{\mathrm{f}}\xspace}

 \def\Pi      {\ensuremath{\mathrm{i}}\xspace}

 \def\Pp      {\ensuremath{\mathrm{p}}\xspace}

 \def\Ps      {\ensuremath{\mathrm{s}}\xspace}

 \def\thebaroffset{0.0em}
}
{

 \def\Pmu         {\ensuremath{\mu}\xspace}

 \def\Ppi         {\ensuremath{\pi}\xspace}                 
                  
 \def\Prho        {\ensuremath{\rho}\xspace}

 \def\Pphi        {\ensuremath{\phi}\xspace}                 
                  
 \def\Pchi        {\ensuremath{\chi}\xspace}                 
 \def\Ppsi        {\ensuremath{\psi}\xspace}                 
                  
 \mathchardef\PDelta="7101
 \mathchardef\PXi="7104
 \mathchardef\PLambda="7103
 \mathchardef\PSigma="7106
 \mathchardef\POmega="710A
 \mathchardef\PUpsilon="7107
 \mathchardef\PPi="7105
                  
 \def\PB      {\ensuremath{B}\xspace}                 
                  
 \def\PD      {\ensuremath{D}\xspace}

 \def\PJ      {\ensuremath{J}\xspace}                 
 \def\PK      {\ensuremath{K}\xspace}

 \def\PS      {\ensuremath{S}\xspace}

 \def\Pb      {\ensuremath{b}\xspace}                 
 \def\Pc      {\ensuremath{c}\xspace}

 \def\Pf      {\ensuremath{f}\xspace}

 \def\Pi      {\ensuremath{i}\xspace}

 \def\Pp      {\ensuremath{p}\xspace}

 \def\Ps      {\ensuremath{s}\xspace}

 \def\thebaroffset{0.18em}
}
\newcommand{\offsetoverline}[2][\thebaroffset]{\kern #1\overline{\kern -#1 #2}}%

\makeatletter
\ifcase \@ptsize \relax
  \newcommand{\miniscule}{\@setfontsize\miniscule{4}{5}}
\or
  \newcommand{\miniscule}{\@setfontsize\miniscule{5}{6}}
\or
  \newcommand{\miniscule}{\@setfontsize\miniscule{5}{6}}
\fi
\makeatother

\DeclareRobustCommand{\optbar}[1]{\shortstack{{\miniscule (\rule[.5ex]{1.25em}{.18mm})}
  \\ [-.7ex] $#1$}}

\def\mumu       {{\ensuremath{\Pmu^+\Pmu^-}}\xspace}

\def\squark    {{\ensuremath{\Ps}}\xspace}

\def\cquark    {{\ensuremath{\Pc}}\xspace}

\def\bquark    {{\ensuremath{\Pb}}\xspace}

\def\pion   {{\ensuremath{\Ppi}}\xspace}

\def\pip    {{\ensuremath{\pion^+}}\xspace}
\def\pim    {{\ensuremath{\pion^-}}\xspace}

\def\kaon    {{\ensuremath{\PK}}\xspace}

\def\KorKbar {\kern \thebaroffset\optbar{\kern -\thebaroffset \PK}{}\xspace}
\def\Kz      {{\ensuremath{\kaon^0}}\xspace}

\def\Kp      {{\ensuremath{\kaon^+}}\xspace}
\def\Km      {{\ensuremath{\kaon^-}}\xspace}

\def\KS      {{\ensuremath{\kaon^0_{\mathrm{S}}}}\xspace}

\def\Dbar    {{\ensuremath{\offsetoverline{\PD}}}\xspace}
\def\D       {{\ensuremath{\PD}}\xspace}

\def\DorDbar {\kern \thebaroffset\optbar{\kern -\thebaroffset \PD}\xspace}
\def\Dz      {{\ensuremath{\D^0}}\xspace}

\def\Dp      {{\ensuremath{\D^+}}\xspace}
\def\Dm      {{\ensuremath{\D^-}}\xspace}

\def\DpDm    {\ensuremath{\Dp {\kern -0.16em \Dm}}\xspace}
\def\Dstar   {{\ensuremath{\D^*}}\xspace}

\def\Dstarzb {{\ensuremath{\Dbar{}^{*0}}}\xspace}

\def\Dstarp  {{\ensuremath{\D^{*+}}}\xspace}

\def\Ds      {{\ensuremath{\D^+_\squark}}\xspace}

\def\B       {{\ensuremath{\PB}}\xspace}

\def\BorBbar {\kern \thebaroffset\optbar{\kern -\thebaroffset \PB}\xspace}

\def\Bd      {{\ensuremath{\B^0}}\xspace}

\def\BdorBdbar {\kern \thebaroffset\optbar{\kern -\thebaroffset \Bd}\xspace}
\def\Bu      {{\ensuremath{\B^+}}\xspace}

\def\Bp      {{\ensuremath{\Bu}}\xspace}

\def\Bs      {{\ensuremath{\B^0_\squark}}\xspace}

\def\BsorBsbar {\kern \thebaroffset\optbar{\kern -\thebaroffset \Bs}\xspace}

\def\jpsi     {{\ensuremath{{\PJ\mskip -3mu/\mskip -2mu\Ppsi}}}\xspace}
\def\psitwos  {{\ensuremath{\Ppsi{(2\PS)}}}\xspace}

\def\chicone  {{\ensuremath{\Pchi_{\cquark 1}}}\xspace}

\def\Y#1S{\ensuremath{\PUpsilon{(#1S)}}\xspace}

\def\proton      {{\ensuremath{\Pp}}\xspace}

\def\Lz          {{\ensuremath{\PLambda}}\xspace}

\def\LorLbar     {\kern \thebaroffset\optbar{\kern -\thebaroffset \PLambda}\xspace}

\def\Lb           {{\ensuremath{\Lz^0_\bquark}}\xspace}

\def\BF         {{\ensuremath{\mathcal{B}}}\xspace}
\def\BR         {\BF}

\newcommand{\decay}[2]{\ensuremath{#1\!\to #2}\xspace} 

\def\to                 {\ensuremath{\rightarrow}\xspace}

\def\CP                {{\ensuremath{C\!P}}\xspace}

\def\AT#1     {\ensuremath{A_{\mathrm{T}}^{#1}}\xspace}           

\def\C#1      {\ensuremath{\mathcal{C}_{#1}}\xspace}                       
\def\Cp#1     {\ensuremath{\mathcal{C}_{#1}^{'}}\xspace}                    
\def\Ceff#1   {\ensuremath{\mathcal{C}_{#1}^{\mathrm{(eff)}}}\xspace}        
\def\Cpeff#1  {\ensuremath{\mathcal{C}_{#1}^{'\mathrm{(eff)}}}\xspace}       
\def\Ope#1    {\ensuremath{\mathcal{O}_{#1}}\xspace}                       
\def\Opep#1   {\ensuremath{\mathcal{O}_{#1}^{'}}\xspace}                    

\newcommand{\nospaceunit}[1]{\ensuremath{\text{#1}}}       
\newcommand{\aunit}[1]{\ensuremath{\text{\,#1}}}       

\newcommand{\tev}{\aunit{Te\kern -0.1em V}\xspace}
\newcommand{\gev}{\aunit{Ge\kern -0.1em V}\xspace}
\newcommand{\mev}{\aunit{Me\kern -0.1em V}\xspace}
\newcommand{\kev}{\aunit{ke\kern -0.1em V}\xspace}
\newcommand{\ev}{\aunit{e\kern -0.1em V}\xspace}
 
\newcommand{\mevc}{\ensuremath{\aunit{Me\kern -0.1em V\!/}c}\xspace}
\newcommand{\gevc}{\ensuremath{\aunit{Ge\kern -0.1em V\!/}c}\xspace}
\newcommand{\mevcc}{\ensuremath{\aunit{Me\kern -0.1em V\!/}c^2}\xspace}
\newcommand{\gevcc}{\ensuremath{\aunit{Ge\kern -0.1em V\!/}c^2}\xspace}

\def\mm   {\aunit{mm}\xspace}

\def\mum  {\ensuremath{\,\upmu\nospaceunit{m}}\xspace}

\def\fb   {\ensuremath{\aunit{fb}}\xspace}
\def\invfb   {\ensuremath{\fb^{-1}}\xspace}

\newcommand{\chisq}{\ensuremath{\chi^2}\xspace}

\def\gsim{{~\raise.15em\hbox{$>$}\kern-.85em
          \lower.35em\hbox{$\sim$}~}\xspace}
\def\lsim{{~\raise.15em\hbox{$<$}\kern-.85em
          \lower.35em\hbox{$\sim$}~}\xspace}

\def\sPlot{\mbox{\em sPlot}\xspace}

\def\pt         {\ensuremath{p_{\mathrm{T}}}\xspace}

\def\evtgen     {\mbox{\textsc{EvtGen}}\xspace}

\def\geant      {\mbox{\textsc{Geant4}}\xspace}

\def\photos     {\mbox{\textsc{Photos}}\xspace}

\def\pythia     {\mbox{\textsc{Pythia}}\xspace}

\def\tell1  {TELL1\xspace}
\def\ukl1   {UKL1\xspace}

\newcommand{\lhcborcid}[1]{\href{https://orcid.org/#1}{\hspace*{0.1em}\raisebox{-0.45ex}{\includegraphics[width=1em]{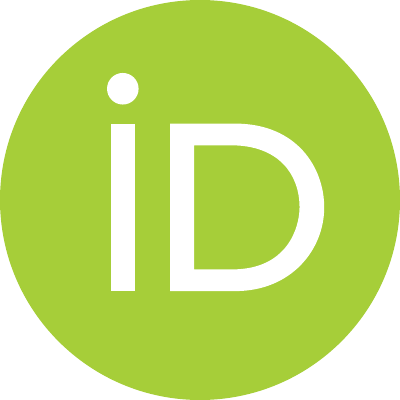}}}}

\usepackage{boldline}
\usepackage{bm}

\usepackage{cite} 
\usepackage{mciteplus}

\begin{document}

\renewcommand{\thefootnote}{\fnsymbol{footnote}}
\setcounter{footnote}{1}


\begin{titlepage}
\pagenumbering{roman}

\vspace*{-1.5cm}
\centerline{\large EUROPEAN ORGANIZATION FOR NUCLEAR RESEARCH (CERN)}
\vspace*{1.5cm}
\noindent
\begin{tabular*}{\linewidth}{lc@{\extracolsep{\fill}}r@{\extracolsep{0pt}}}
\ifthenelse{\boolean{pdflatex}}
{\vspace*{-1.5cm}\mbox{\!\!\!\includegraphics[width=.14\textwidth]{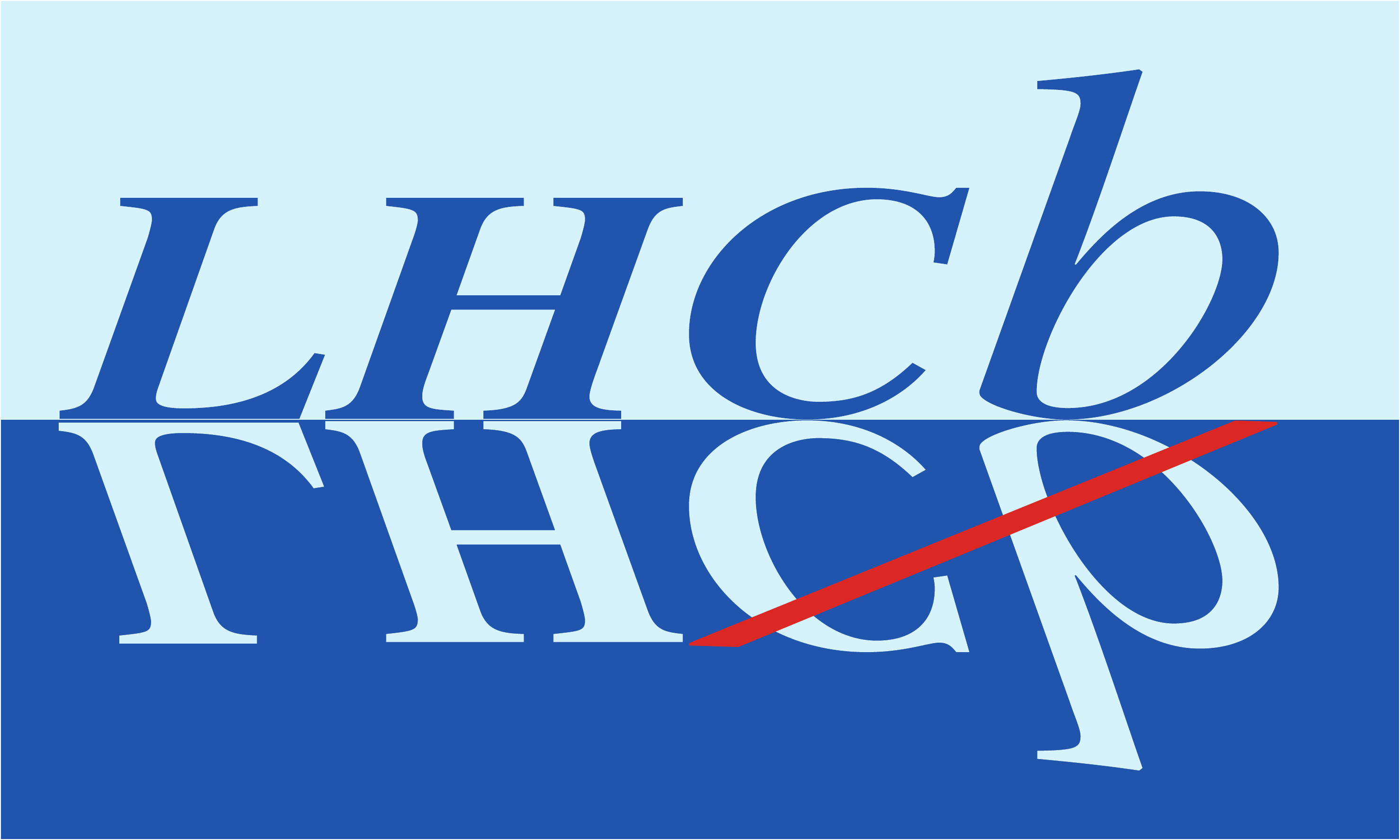}} & &}%
{\vspace*{-1.2cm}\mbox{\!\!\!\includegraphics[width=.12\textwidth]{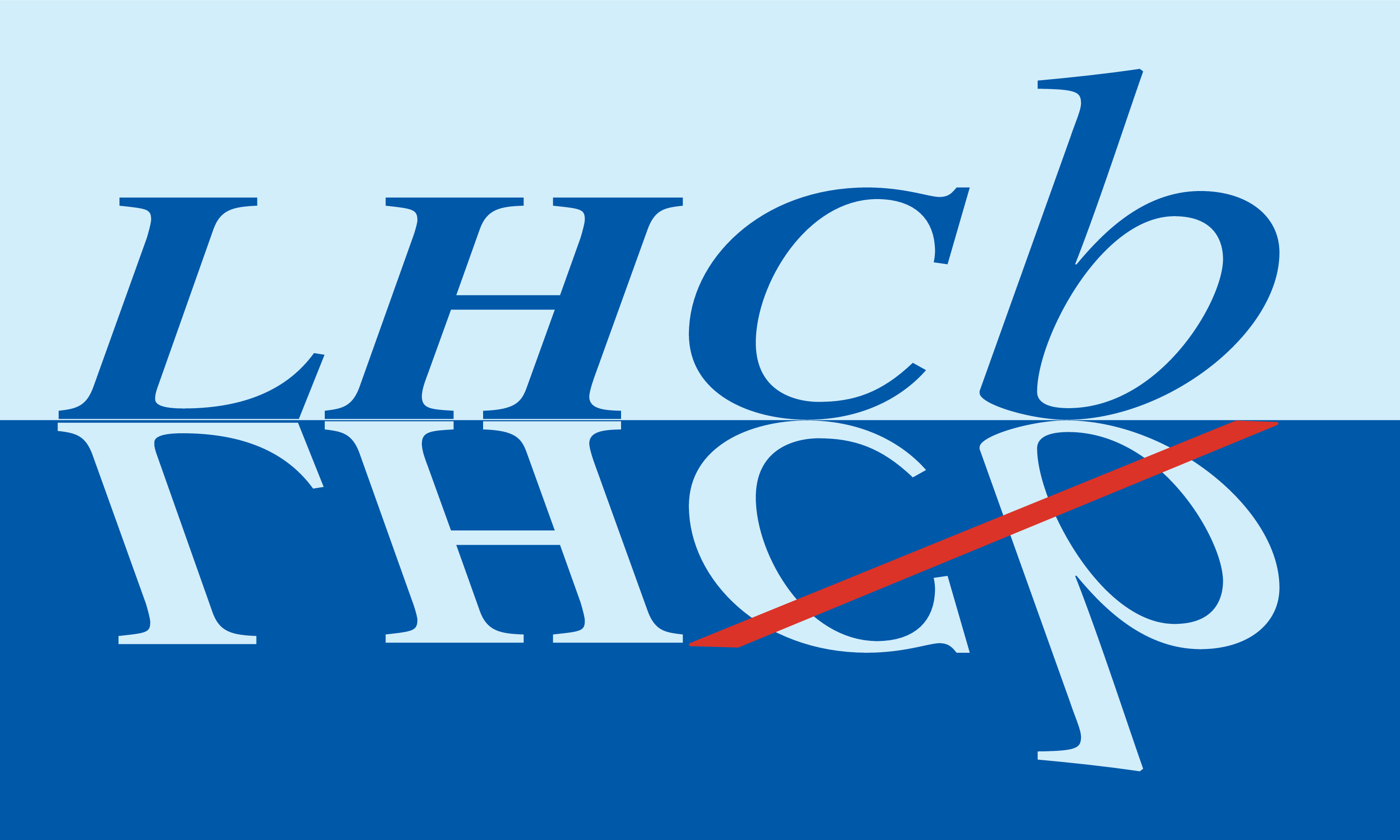}} & &}%
\\
 & & CERN-EP-2023-015 \\  
 & & LHCb-PAPER-2022-049 \\  
 & & February 18, 2023\\ 
 & & \\
\end{tabular*}

\vspace*{4.0cm}

{\normalfont\bfseries\boldmath\huge
\begin{center}
  \papertitle 
\end{center}
}

\vspace*{2.0cm}

\begin{center}
\paperauthors\footnote{Authors are listed at the end of this paper.}
\end{center}

\vspace{\fill}

\begin{abstract}
\noindent
The~first observation of 
the~\mbox{$\decay{\Bs}{\left( \decay{\chicone(3872)}{\jpsi\pip\pim}\right) 
\pip \pim}$}~decay is reported 
using 
proton\nobreakdash-proton 
collision  data, 
corresponding to 
integrated luminosities of 1, 2 and 6\invfb, 
collected by the~LHCb experiment
at centre\nobreakdash-of\nobreakdash-mass
energies of 7, 8 and 13\tev, respectively. 
The~ratio of branching fractions relative to 
the~\mbox{$\decay{\Bs}{\left( \decay{\psitwos}{\jpsi\pip\pim}
\right) \pip \pim}$} 
decay is measured to be 
$$
\dfrac{ \BR\left(\decay{\Bs}
{ \chicone(3872)  \pip\pim}\right) 
\times \BR\left(\decay{\chicone(3872)}{\jpsi\pip\pim}\right)}
{\BR\left(\decay{\Bs}
{ \psitwos  \pip\pim}\right) 
\times \BR\left(\decay{\psitwos}{\jpsi\pip\pim}\right)}
 = \left( 6.8  \pm 1.1 \pm 0.2 \right) \times 10^{-2}\,,
$$ 
where the first uncertainty is statistical and 
the~second systematic. 
The~mass spectrum of the~$\pip\pim$~system 
recoiling against the~$\chicone(3872)$~meson 
exhibits a~large contribution from 
\mbox{$\decay{\Bs}{\chicone(3872)\left(\decay{\Pf_0(980)} 
{\pip \pim} \right)}$}~decays. 

\end{abstract}

\vspace*{2.0cm}

\begin{center}
  Published in \href{https://doi.org/10.1007/JHEP07(2023)084}{JHEP 07 (2023) 084}
\end{center}

\vspace{\fill}

{\footnotesize 
\centerline{\copyright~\papercopyright. \href{\paperlicenceurl}{\paperlicence}.}}
\vspace*{2mm}

\end{titlepage}


\newpage
\setcounter{page}{2}
\mbox{~}
%
%
%
%


\renewcommand{\thefootnote}{\arabic{footnote}}
\setcounter{footnote}{0}

\cleardoublepage


\pagestyle{plain} 
\setcounter{page}{1}
\pagenumbering{arabic}


\section{Introduction}
\label{sec:Introduction}

Decays of beauty hadrons to final states with charmonia
provide a~unique laboratory to study 
the~properties of charmonia and 
charmonium\nobreakdash-like states. 
A~plethora of new states has been observed in such 
decays, 
including the~$\chicone(3872)$~particle~\cite{Choi:2003ue},
pentaquark~\mbox{\cite{LHCb-PAPER-2015-029,
LHCb-PAPER-2016-009,
LHCb-PAPER-2019-014,
LHCb-PAPER-2016-015,
LHCb-PAPER-2020-039}}
and 
tetraquark~\cite{
Choi:2007wga,
Mizuk:2009da,
Chilikin:2013tch,
LHCb-PAPER-2014-014,
LHCb-PAPER-2015-038,
LHCb-PAPER-2016-018,
LHCb-PAPER-2016-019,
LHCb-PAPER-2018-034,
LHCb-PAPER-2018-043,
LHCb-PAPER-2020-035,
LHCb-PAPER-2020-044,
LHCb-PAPER-2021-018}
candidates 
as well as conventional charmonium states, such as 
the~tensor D\nobreakdash-wave 
$\Ppsi_2(3823)$~meson~\cite{Bhardwaj:2013rmw,
LHCb-PAPER-2020-009,LHCb-PAPER-2021-047}. 
The~nature of many exotic 
charmonium\nobreakdash-like  candidates 
remains unclear. 
A~comparison of their production rates 
with respect to those  
of conventional  charmonium states 
in decays 
of beauty hadrons can shed light on 
their~production mechanisms~\cite{Maiani:2017kyi}.
For~example, 
the~$\Dstar\Dbar$~rescattering 
mechanism~\cite{Artoisenet:2010va,Braaten:2019yua}
would give a~large contribution 
to $\chicone(3872)$~production
and affect the~pattern of decay 
rates of beauty hadrons. 
There~is 
a~puzzling difference between 
the~branching fractions for 
the~\mbox{$\decay{\Bu}{\chicone(3872)\Kp}$}
and~\mbox{$\decay{\Bd}{\chicone(3872)\Kz}$}~decays~\cite{PDG2022,
Belle:2023zxm}.
It~may be explained by 
a~compact\nobreakdash-tetraquark interpretation
of
the~\mbox{$\chicone(3872)$}~state~\cite{Maiani:2020zhr},
which 
simultaneously accounts for 
the~similarity of 
the~branching fractions for
the~\mbox{$\decay{\Bd}{\chicone(3872)\Kz}$}
and 
\mbox{$\decay{\Bs}{\chicone(3872)\Pphi}$}~decays~\cite{Sirunyan:2020qir,
LHCb-PAPER-2020-035}. 
Additional measurements on the~$\chicone(3872)$~production 
in the~decays of beauty hadrons,
and in particular, decays of \Bs~mesons, 
will be helpful for a~better understanding 
of the~nature of the~\mbox{$\chicone(3872)$}~state.

In this paper, 
the~first observation of 
the~\mbox{$\decay{\Bs}{\chicone(3872)\pip\pim}$}~decay
is  reported.
This analysis is based on 
proton\nobreakdash-proton\,($\proton\proton$) collision  data, 
corresponding to 
integrated luminosities of 1, 2 and 6\invfb, 
collected by the~LHCb experiment
at centre\nobreakdash-of\nobreakdash-mass
energies of 7, 8 and 13\tev, respectively. 
    The~\mbox{$\decay{\Bs}{\psitwos\pip\pim}$}~decay
    is used as normalisation channel  
    to measure the~ratio $\mathcal{R}$ of 
the~branching fractions 
of the~\mbox{$\decay{\Bs}{ \left( \decay{\chicone(3872)}{\jpsi\pip\pim} \right)  \pip\pim}$} 
and  \mbox{$\decay{\Bs}{ \left( \decay{\psitwos}{\jpsi\pip\pim} 
    \right) \pip\pim  }$}~decays
\begin{equation}\label{eq:r}
\mathcal{R} 
\equiv    
\dfrac{ \BR\left(\decay{\Bs}
{ \chicone(3872)  \pip\pim}\right) 
\times \BR\left(\decay{\chicone(3872)}{\jpsi\pip\pim}\right)}
{\BR\left(\decay{\Bs}
{ \psitwos  \pip\pim}\right) 
\times \BR\left(\decay{\psitwos}{\jpsi\pip\pim}\right)}\,. 
\end{equation}
The~\mbox{$\decay{\Bs}
{ \left( \decay{\chicone(3872)}
{\jpsi\pip\pim} \right)  \pip\pim}$} 
and  \mbox{$\decay{\Bs}{ \left( \decay{\psitwos}
{\jpsi\pip\pim} 
\right) \pip\pim  }$}~decays
share the~same final state, 
allowing for a~large cancellation of
systematic uncertainties.

\section{Detector and simulation}
\label{sec:Detector}

The \lhcb detector~\cite{LHCb-DP-2008-001,LHCb-DP-2014-002} is a single-arm forward
spectrometer covering the~pseudorapidity range $2<\eta <5$,
designed for the study of particles containing $\bquark$~or~$\cquark$~quarks. 
The~detector includes a high-precision tracking system consisting of a 
silicon-strip vertex detector surrounding the \proton\proton interaction
region~\cite{LHCb-DP-2014-001}, a large-area silicon-strip detector located
upstream of a dipole magnet with a bending power of about $4{\mathrm{\,Tm}}$,
and three stations of silicon-strip detectors and straw drift tubes~\cite{LHCb-DP-2013-003,LHCb-DP-2017-001} placed downstream of the magnet. 
The tracking system provides a measurement of the momentum of charged particles
with a relative uncertainty that varies from $0.5\%$ at low momentum to $1.0\%$~at~$200 \gevc$. 
The~momentum scale is calibrated using samples of $\decay{\jpsi}{\mumu}$ 
and $\decay{\Bu}{\jpsi\Kp}$~decays collected concurrently
with the~data sample used for this analysis~\cite{LHCb-PAPER-2012-048,LHCb-PAPER-2013-011}. 
The~relative accuracy of this
procedure is estimated to be $3 \times 10^{-4}$ using samples of other
fully reconstructed $\bquark$~hadrons, $\PUpsilon$~and
$\KS$~mesons.
The~minimum distance of a track to a primary $\proton\proton$\nobreakdash-collision vertex\,(PV), 
the~impact parameter\,(IP), 
is~measured with a~resolution of $(15+29/\pt)\mum$, where \pt is the component 
of the~momentum transverse to the beam, in\,\gevc. Different types of charged hadrons
are distinguished using information from two ring-imaging Cherenkov detectors\,(RICH)~\cite{LHCb-DP-2012-003}. Photons,~electrons and hadrons are identified 
by a~calorimeter system consisting of scintillating\nobreakdash-pad 
and preshower detectors, 
an electromagnetic and 
a~hadronic calorimeter~\cite{LHCb-DP-2020-001}. Muons are~identified by a~system 
composed of alternating layers of iron and multiwire proportional chambers~\cite{LHCb-DP-2012-002}.

The online event selection is performed by a trigger~\cite{LHCb-DP-2012-004}, 
which consists of a hardware stage, based on information from the calorimeter and muon systems,
followed by a~software stage, which applies a~full event reconstruction. 
The hardware trigger selects muon candidates with large 
transverse momentum or dimuon candidates with a~large value of 
the~product
of the~$\pt$ of the~muons. 
In~the~software trigger, two 
oppositely charged muons are required to form 
a~good\nobreakdash-quality
vertex that is significantly displaced from every~PV,
with a~dimuon mass exceeding~$2.7\gevcc$.

Simulated events are used to describe signal  
shapes
and to~compute the efficiencies needed to determine 
the~branching fraction ratio.
In~the~simulation, \proton\proton collisions 
are generated 
using \pythia~\cite{Sjostrand:2007gs}  
with a~specific \lhcb configuration~\cite{LHCb-PROC-2010-056}. 
Decays of unstable particles are described by 
the~\evtgen 
package~\cite{Lange:2001uf}, 
in which final\nobreakdash-state radiation is generated 
using \photos~\cite{Golonka:2005pn}. 
The~interaction of the~generated particles with the~detector, 
and its response, are implemented using
the~\geant toolkit~\cite{Allison:2006ve, *Agostinelli:2002hh} 
as described in Ref.~\cite{LHCb-PROC-2011-006}.
The~decays 
\mbox{$\decay{\Bs}{ \chicone(3872)\pip\pim}$} 
and 
\mbox{$\decay{\Bs}{ \psitwos\pip\pip}$} 
are simulated using 
a~phase-space decay model
that is 
adjusted to match 
the~mass distributions 
of the~two\nobreakdash-pion systems 
recoiling against the~$\chicone(3872)$ and $\psitwos$~mesons in data. 
In~the~simulation
\mbox{$\decay{\chicone(3872)}
{\jpsi\pip\pim}$}~decays
proceed
via an~S\nobreakdash-wave $\jpsi\Prho^0$ 
intermediate state~\cite{LHCb-PAPER-2013-001,
LHCb-PAPER-2015-015,
LHCb-PAPER-2021-045}. 
The~model described in 
Refs.~\cite{Gottfried:1977gp,
Voloshin:1978hc,
Peskin:1979va,
Bhanot:1979vb,
Voloshin:1980zf,
Novikov:1980fa}
is used for 
the~\mbox{$\decay{\psitwos}{\jpsi\pip\pim}$} decays.
The~simulation is corrected to reproduce 
the~transverse momentum and 
rapidity distributions of the~$\Bs$~mesons 
observed in data.
To~account for imperfections in the~simulation of
charged\nobreakdash-particle reconstruction, 
the~track reconstruction efficiency
determined from simulation 
is corrected using data\nobreakdash-driven
techniques~\cite{LHCb-DP-2013-002}.

\section{Event selection}
\label{sec:evt_sel}

Candidate 
$\decay{\Bs}{\jpsi\pip\pip\pim\pim}$~decays
are reconstructed 
using the~$\decay{\jpsi}{\mumu}$~decay mode.
As explained in detail below,
an~initial selection criteria similar  
to those used 
in Refs.~\cite{LHCb-PAPER-2015-060,
LHCb-PAPER-2019-023,
LHCb-PAPER-2020-035}
are used to reduce
the~background. 
Subsequently, 
a~multivariate estimator,
in the~following referred as 
the~{\sc{MLP}} classifier, 
is applied.
It is based on an~artificial neural 
network algorithm~\cite{McCulloch,rosenblatt58} 
configured with a~cross\nobreakdash-entropy cost 
estimator~\cite{Zhong:2011xm}.

Muon and hadron candidates are identified 
using combined information from 
the~RICH, calorimeter and muon 
detectors~\cite{LHCb-PROC-2011-008}. 
The~candidates  are required to have a~transverse 
momentum greater 
than $550\mevc$ and $200\mevc$ 
for muons and pions, respectively. 
To~ensure that
the~particles can be efficiently 
separated by the~RICH detectors,
pions are required to have a~momentum 
between $3.2$~and $150\gevc$. 
To~reduce the~combinatorial background 
due to particles 
produced promptly
in the~$\proton\proton$~interaction, 
only tracks that are inconsistent with 
originating 
from  a~PV are used. 
Pairs of oppositely charged muons consistent with originating 
from a~common vertex are combined to form $\jpsi$ candidates. 
The~mass of the~dimuon candidate is required 
to be between $3.05$ and $3.15\gevcc$, 
corresponding 
to 
a~range of 
approximately 
three times  
the~\mumu~mass resolution, around 
the~known mass of 
the~\jpsi~meson~\cite{PDG2022}.
Selected $\jpsi$~meson candidates are 
combined with two pairs of oppositely charged pions 
to form the~$\decay{\Bs}{\jpsi\pip\pip\pim\pim}$~candidates
and a~requirement on the~quality 
of the~common six\nobreakdash-prong vertex is imposed.  
To~improve the~mass and decay time resolution, 
a~kinematic fit~\cite{Hulsbergen:2005pu} is used 
in which 
the~momentum 
direction of the~\Bs~candidate
is constrained 
to be collinear to 
the~direction from the~PV to 
the~\Bs~decay vertex, 
and 
a~mass constraint on the~\jpsi~state is applied. 
A~requirement on the~\chisq of this fit, 
$\chisq_{\mathrm{fit}}$, 
is imposed to reduce the~background.
The~mass of selected $\jpsi\pip\pip\pim\pim$~combinations,
$m_{\jpsi\pip\pip\pim\pim}$, 
is required to be between 
$5.30$ and $5.48\gevcc$.
The~proper decay time of the~$\Bs$~candidates 
is required to be 
between 0.2 and $2.0\mm/c$. 
The~lower limit is used to reduce 
background from particles coming 
from the~PV
while the~upper limit suppresses poorly 
reconstructed candidates.
A~possible feed down 
from~\mbox{$\decay{\Lb}{\jpsi\proton\pip\pim\pim}$}~decays,  
with the~proton misidentified as a~pion, 
is suppressed by rejecting the~\Bs~candidates
whose mass, recalculated using 
the~proton hypothesis  for 
one of the~pion candidates, 
is consistent with the~known mass  
of the~\Lb~baryon~\cite{PDG2022}.

The~final selection of candidates using 
the~{\sc{MLP}}~classifier is based on 
the \pt and pseudorapidity of \jpsi~candidate,
the \chisq~of the~six\nobreakdash-prong vertex,
the value of $\chisq_{\mathrm{fit}}$,  
the~proper decay time of the~\Bs~candidates, 
the transverse momenta of the pions, 
the~dipion mass 
from the~\mbox{$\decay{\chicone(3872)}
{\jpsi\pip\pim}$}~decays
and  the~angle between 
the~momenta of 
the~$\jpsi$ and \Bs~mesons  
in the~$\chicone(3872)$~rest frame. 
The~{\sc{MLP}}~classifier is trained on 
a~sample
of simulated \mbox{$\decay{\Bs}{\left(\decay{\chicone(3872)}
{\jpsi\pip\pim}\right)\pip\pim}$}~decays
and a~background sample
of \mbox{$\jpsi\pip\pip\pim\pim$}~combinations 
from the~high mass sideband of the~\Bs~signal peak,
\mbox{$5.42<m_{\jpsi\pip\pip\pim\pim}<5.50\gevcc$}.
         The~low\nobreakdash-mass
         sideband of the~\Bs~signal peak contains
         a~large contribution from partially reconstructed
         \bquark-hadron decays and therefore is not representative
         of the~background under the~\Bs~signal peak. 
For the~background sample, 
$\jpsi\pip\pim$~combinations 
with the~$\jpsi\pip\pim$~mass 
consistent 
with the~known masses 
of 
the~$\chicone(3872)$~state~\cite{LHCb-PAPER-2020-008,
LHCb-PAPER-2020-009},
\mbox{$3.86<m_{\jpsi\pip\pim}<3.88\gevcc$}, 
or \psitwos~meson~\cite{PDG2022},
\mbox{$3.68<m_{\jpsi\pip\pim}<3.69\gevcc$}, 
are excluded. 
To~avoid introducing 
a~bias in the~{\sc{MLP}}~evaluation, 
a~$k$\nobreakdash-fold 
cross\nobreakdash-validation technique~\cite{chopping}
with $k=7$ is used.
The~requirement on the~{\sc{MLP}}~classifier 
is chosen to maximize 
the~Punzi figure of merit
$\tfrac{\varepsilon}
{\upalpha/2+\sqrt{B}}$~\cite{Punzi:2003bu},
where $\varepsilon$ is the~efficiency 
for the~$\decay{\Bs}{\chicone(3872)\pip\pim}$~signal, 
$\upalpha=5$ is the~target signal significance
and $B$ is the~expected background yield.
The~efficiency $\varepsilon$ is 
estimated using simulation. 
The~expected background yield within 
the~narrow mass window centred
around the~known masses of 
the~$\Bs$ and $\chicone(3872)$~mesons~\cite{PDG2022}
is determined from fits to data.

The~selected 
\Bs~candidates 
with $\jpsi\pip\pim$~mass within 
the~range 
\mbox{$3.85<m_{\jpsi\pip\pim}<3.90\gevcc$}
are considered as  
the~\mbox{$\decay{\Bs}{\chicone(3872)\pip\pim}$}~candidates.
Similarly, the~\Bs~candidates with 
the~$\jpsi\pip\pim$~mass within 
the~range 
\mbox{$3.67<m_{\jpsi\pip\pim}<3.70\gevcc$}
are considered as 
the~\mbox{$\decay{\Bs}
{\psitwos\pip\pim}$}~candidates. 
The~same {\sc{MLP}}~classifier is used 
both for 
the~\mbox{$\decay{\Bs}{\chicone(3872)\pip\pim}$}
and \mbox{$\decay{\Bs}{\psitwos\pip\pim}$}~candidates.

\section{
$\decay{\Bs}{\chicone(3872)\pip\pim}$
and $\decay{\Bs}{\psitwos\pip\pim}$~decays}
\label{sec:xcc_pi}

The signal yields for 
the~\mbox{$\decay{\Bs}{\chicone(3872)\pip\pim}$}
and 
\mbox{$\decay{\Bs}{\psitwos\pip\pim}$}~decays 
are determined using 
a~two-dimensional 
extended 
unbinned 
maximum\nobreakdash-likelihood fit 
to the~$\jpsi\pip\pip\pim\pim$ and
$\jpsi\pip\pim$~mass distributions.
    Following Refs.~\cite{LHCb-PAPER-2020-009,
      LHCb-PAPER-2020-035,
      LHCb-PAPER-2022-025},
    to improve the~resolution on
    the~$\jpsi\pip\pim$~mass and
    to eliminate a~small correlation between
    the~$\jpsi\pip\pim$ and
    $\jpsi\pip\pip\pim\pim$~masses,
    the~$\jpsi\pip\pim$~mass is computed
    constraining 
    the~mass of
    the~\Bs~candidate~\cite{Hulsbergen:2005pu}
    to its known mass~\cite{PDG2022}.
The~fit is performed simultaneously 
in
two separate regions of the~$\jpsi\pip\pim$~mass,
defined as 
the~$\chicone(3872)$~region 
with 
\mbox{$3.85<m_{\jpsi\pip\pim}<3.90\gevcc$}, 
and the~$\psitwos$~region with 
\mbox{$3.67<m_{\jpsi\pip\pim}<3.70\gevcc$},
which  
correspond 
to the~\mbox{$\decay{\Bs}
{\chicone(3872)\pip\pim}$}
and~\mbox{$\decay{\Bs}
{\psitwos\pip\pim}$}~decays, respectively.

In~the~$\chicone(3872)$~region, 
the~two\nobreakdash-dimensional fit model is defined 
as the sum  of four components:
\begin{enumerate}
\item A~signal~component, 
corresponding to 
the~\mbox{$\decay{\Bs}
{\left(\decay{\chicone(3872)}
{\jpsi\pip\pim}\right)\pip\pim}$}~decay, 
described by the product of~the~\Bs and  
$\chicone(3872)$~signal templates, 
discussed in the~next paragraph.
\item 
A~component  corresponding 
to~\mbox{$\decay{\Bs}{\jpsi\pip\pip\pim\pim}$}~decay,
where the~$\jpsi\pip\pim$~combination does not originate from
a~$\chicone(3872)$~meson, 
parameterised by 
the~product of 
the~\Bs~signal template 
and the~phase\nobreakdash-space function 
describing three\nobreakdash-body combinations 
from five-body decays\footnote{
The~phase\nobreakdash-space mass distribution of 
a~$k$\nobreakdash-body combination of
particles from an~$n$\nobreakdash-body decay 
is approximated by 
$\Phi_{k,n}(x) \propto x_{\ast}^{ (3k-5)/2}\left(1-x_{\ast}\right)^{3(n-k)/2-1}$,
where 
$x_{\ast}\equiv ( x-x_{\mathrm{min}} ) /(x_{\mathrm{max}}-x_{\mathrm{min}})$, 
and 
$x_{\mathrm{min}}$, $x_{\mathrm{max}}$ denote 
the~minimal   and maximal values of $x$, 
respectively~\cite{Byckling}.
}
$\Phi_{3,5}(m_{\jpsi\pip\pim})$~\cite{Byckling}, 
modified by a~positive linear polynomial function. 
\item A~component corresponding to random 
 combinations of 
 the~$\chicone(3872)$~state 
 with a~$\pip\pim$~pair,
 parameterised as a~product of 
 the~signal $\chicone(3872)$~template 
 and a~second\nobreakdash-order 
 positive\nobreakdash-definite 
 polynomial function~\cite{karlin1953geometry}.

\item A~component corresponding to random 
 $\jpsi\pip\pip\pim\pim$~combinations, 
 parameterized as 
 a~two\nobreakdash-dimensional
 non\nobreakdash-factorizable 
 positive polynomial 
 function~$P_{\mathrm{bkg}}$, 
 that is linear in each variable for 
 a fixed value of the other variable.
\end{enumerate} 
In~the~$\psitwos$~region, 
the~two\nobreakdash-dimensional 
fit model is defined in 
an~equivalent 
way with 
replacement of the~signal $\chicone(3872)$ template 
with the~signal $\psitwos$~template. 
In~this approach we neglect
possible  interference effects between
    the~\mbox{\decay{\Bs}{\chicone(3872)\pip\pim}}
    and  non-resonant \mbox{\decay{\Bs}{\jpsi\pip\pip\pim\pim}}~contributions.

\begin{figure}[t]
  \setlength{\unitlength}{1mm}
  \centering
  \begin{picture}(160,125)
    %
    \put(  0,  63){ 
      \includegraphics*[width=80mm,
      ]{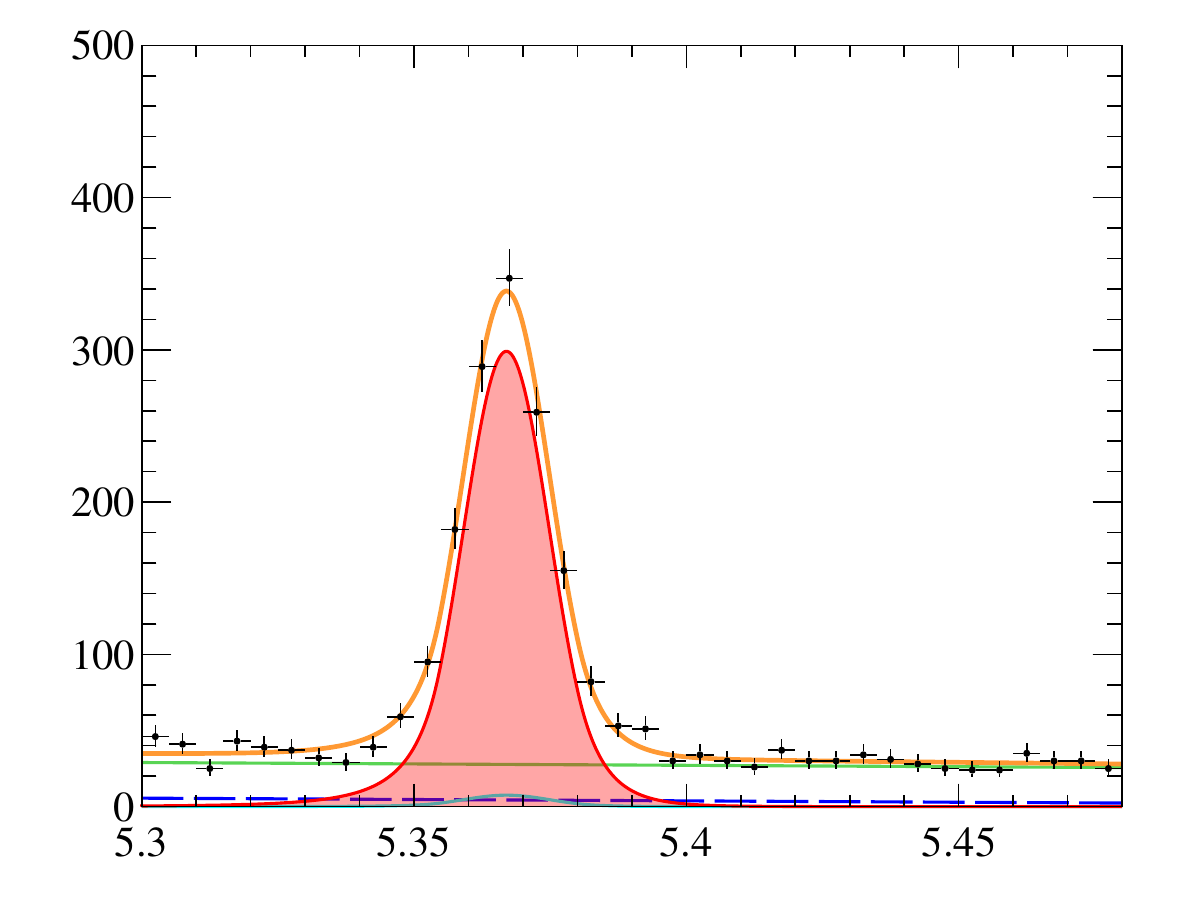}
    }
     \put( 80, 63){ 
      \includegraphics*[width=80mm,
      ]{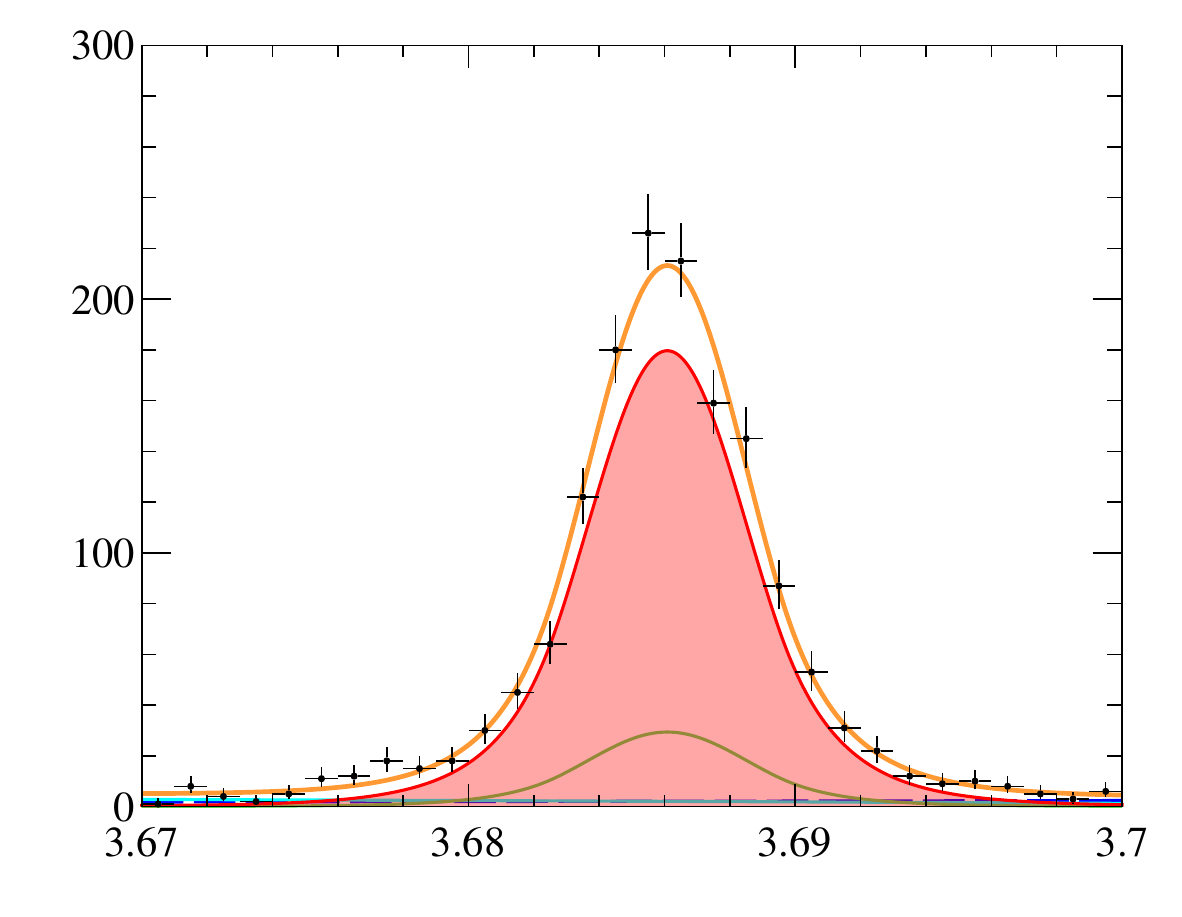}
    }
    \put(  0,  0){ 
      \includegraphics*[width=80mm,
      ]{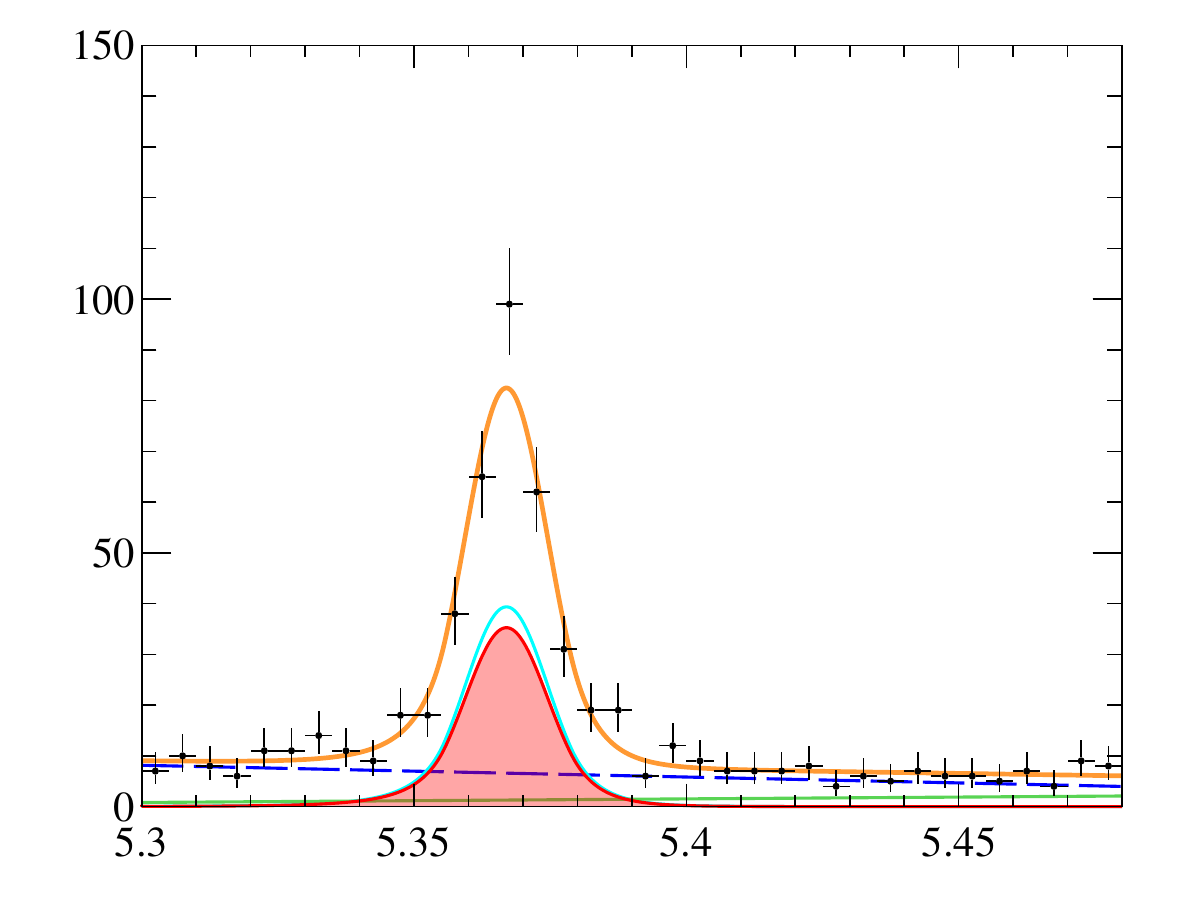}
    }
    \put(  80,  0){ 
      \includegraphics*[width=80mm,
      ]{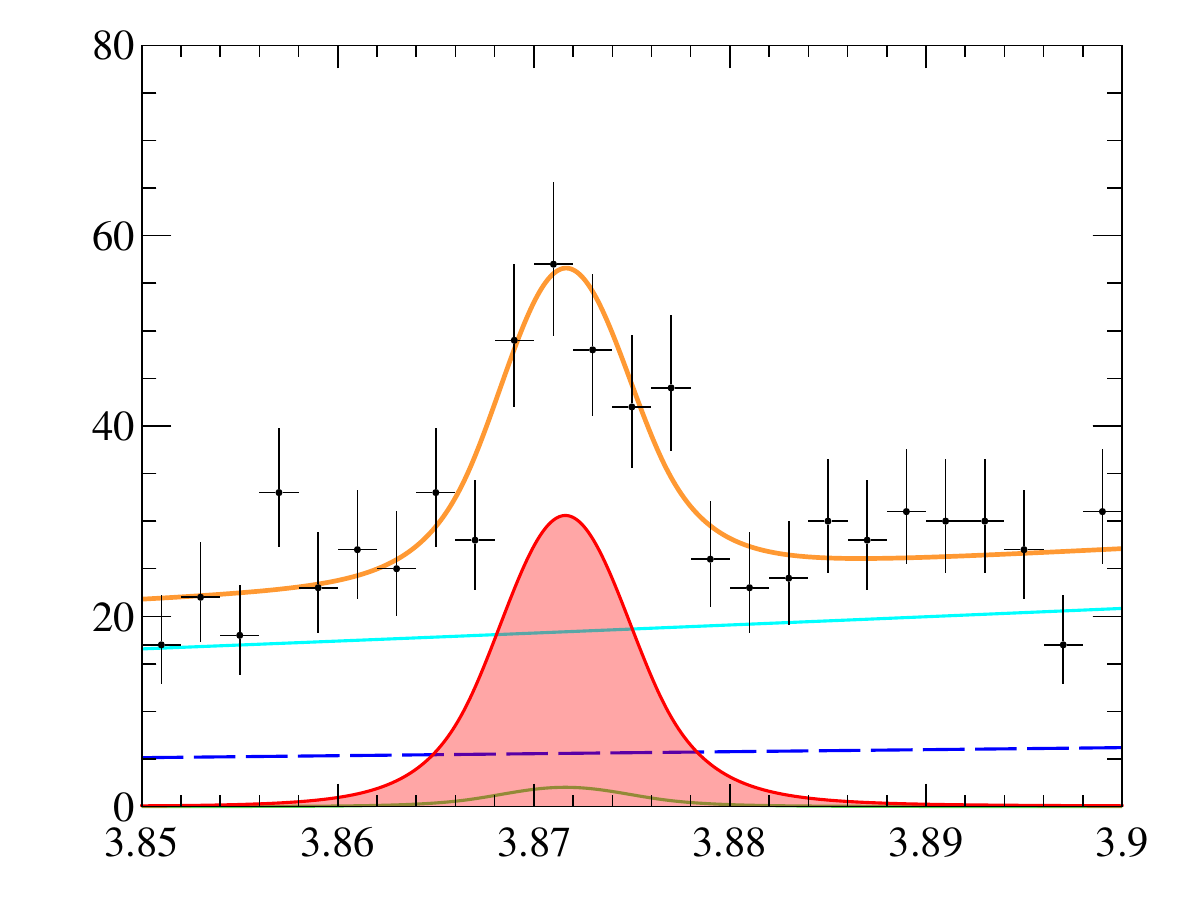}
    }
    \put(   1,  100) {\rotatebox[]{90} 
    {Candidates\,$/\,(5\mevcc)$} }
    \put(  81,  100) {\rotatebox[]{90}
    {Candidates\,$/\,(1\mevcc)$} }
    \put(   1,  37) {\rotatebox[]{90}
    {Candidates$\,/\,(5\mevcc)$} }
    \put(  81,  37) {\rotatebox[]{90}
    {Candidates\,$/\,(2\mevcc)$} }
    \put(  29, 63) {$m_{\jpsi\pip\pip\pim\pim}$} 
    \put(  63, 63) {$\left[\!\gevcc\right]$} 
    \put( 115, 63) {$m_{\jpsi\pip\pim}$} 
    \put( 143, 63) {$\left[\!\gevcc\right]$} 
    \put(  29,  0) {$m_{\jpsi\pip\pip\pim\pim}$} 
    \put(  63,  0) {$\left[\!\gevcc\right]$} 
    \put( 115,  0) {$m_{\jpsi\pip\pim}$} 
    \put( 143,  0) {$\left[\!\gevcc\right]$} 
    \put( 14, 48){$\begin{array}{l}\lhcb\\ 9\invfb\end{array}$}
    \put( 14,111){$\begin{array}{l}\lhcb\\ 9\invfb\end{array}$}
    \put(141, 48){$\begin{array}{l}\lhcb\\ 9\invfb\end{array}$}
    \put(141,111){$\begin{array}{l}\lhcb\\ 9\invfb\end{array}$}
    \definecolor{gr}{rgb}{0.35, 0.83, 0.33}
    \definecolor{br}{rgb}{0.43, 0.98, 0.98}
    \definecolor{vi}{rgb}{0.39, 0.37, 0.96}
    \definecolor{db}{rgb}{0.1, 0.08, 0.41}
    \definecolor{vi}{rgb}{0.39, 0.37, 0.96}
   \put(37,45){\scriptsize
   $\begin{array}{cl}
    \!\bigplus\mkern-5mu & \text{data} 
   \\
   \begin{tikzpicture}[x=1mm,y=1mm]\filldraw[fill=red!35!white,draw=red,thick]  (0,0) rectangle (5,3);\end{tikzpicture}  
   & \decay{\Bs}{\chicone(3872)\pip\pim} 
   \\ 
   {\color[RGB] {110,251,251}{\rule{5mm}{2.0pt}}} 
   & \decay{\Bs}{\jpsi\pip\pip\pim\pim}
   \\ 
   {\color[RGB] {121,220,117}{\rule{5mm}{2.0pt}}}
   & \text{comb.}\,\chicone(3872)\pip\pim
   \\
   {\color[RGB]{85,83,246}{\hdashrule[0.0ex][x]{5mm}{1.0pt}{2.0mm 0.3mm}}}
   & \text{comb.}\,\jpsi\pip\pip\pim\pim
   \\ 
   {\color[RGB]{255,153,51} {\rule{5mm}{2.0pt}}}
   & \text{total}
   \end{array}$ }
   \put(37,108){\scriptsize
   $\begin{array}{cl}
    \!\bigplus\mkern-5mu & \text{data} 
   \\
   \begin{tikzpicture}[x=1mm,y=1mm]\filldraw[fill=red!35!white,draw=red,thick]  (0,0) rectangle (5,3);\end{tikzpicture}  
   & \decay{\Bs}{\psitwos\pip\pim} 
   \\ 
   {\color[RGB]{110,251,251} {\rule{5mm}{2.0pt}}} 
   & \decay{\Bs}{\jpsi\pip\pip\pim\pim}
   \\ 
   {\color[RGB] {121,220,117}  {\rule{5mm}{2.0pt}}}
   & \text{comb.}\,\psitwos\pip\pim
   \\
   {\color[RGB]{85,83,246}{\hdashrule[0.0ex][x]{5mm}{1.0pt}{2.0mm 0.3mm}}}
   & \text{comb.}\,\jpsi\pip\pip\pim\pim
   \\ 
   {\color[RGB]{255,153,51} {\rule{5mm}{2.0pt}}}
   & \text{total}
   \end{array}$ }
  \end{picture}
  \caption { \small
   Distributions of 
   the~(left)~\mbox{$\jpsi\pip\pip\pim\pim$}
   and 
   (right)~\mbox{$\jpsi\pip\pim$}~mass
   of selected 
   (top) \mbox{$\decay{\Bs}{\psitwos\pip\pim}$}
   and 
   (bottom) \mbox{$\decay{\Bs}{\chicone(3872)\pip\pim}$}~candidates.
   Projections from the~fit, 
   described in the~text, are overlaid. 
    The~\mbox{$\jpsi\pip\pip\pim\pim$}~mass distributions 
   are shown for
   the~\mbox{$\decay{\Bs}{\psitwos\pip\pim}$}
   and 
   \mbox{$\decay{\Bs}{\chicone(3872)\pip\pim}$}~candidates
   within
   narrow \mbox{$\jpsi\pip\pim$}~mass ranges,
   \mbox{$3.679<m_{\jpsi\pip\pim}<3.693\gevcc$} and 
   \mbox{$3.864<m_{\jpsi\pip\pim}<3.880\gevcc$}, respectively.
   The~\mbox{$\jpsi\pip\pim$}~mass distributions 
   are shown for 
   the~\Bs~candidates within 
   a~narrow 
   \mbox{$\jpsi\pip\pip\pim\pim$}~mass range,
   \mbox{$5.35<m_{\jpsi\pip\pip\pim\pim}<5.38\gevcc$}. 
  }
  \label{fig:fit_sim_chi}
\end{figure}

The~$\Bs$ signal shape is modelled 
with a~modified 
Gaussian function with power\nobreakdash-law 
tails 
on both sides of 
the~distribution~\cite{LHCb-PAPER-2011-013,
Skwarnicki:1986xj}.
The~tail parameters are fixed from~simulation, 
while the~mass parameter of the~\Bs~meson is 
allowed to vary.
The~detector resolution taken 
from simulation 
is corrected 
by a~scale factor, $s_{\Bs}$, 
that accounts for a~small discrepancy 
between data and simulation~\cite{LHCb-PAPER-2020-009,
LHCb-PAPER-2020-035}
and 
is allowed to vary in the~fit.

The~$\chicone(3872)$ and $\psitwos$~signal 
shapes are modelled 
as relativistic S\nobreakdash-wave 
Breit\nobreakdash--Wigner
functions
convolved with the~detector 
resolution.
Due~to the~proximity of 
the~$\chicone(3872)$~state to 
the~$\Dz\Dstarzb$~threshold,
modelling this component 
with a~Breit\nobreakdash--Wigner 
function may not be adequate~\cite{Hanhart:2007yq,
Stapleton:2009ey,
Kalashnikova:2009gt,
Artoisenet:2010va,
Hanhart:2011jz}. 
However, the~analyses in Refs.~\cite{LHCb-PAPER-2020-008,
LHCb-PAPER-2020-009}
demonstrate that a~good description of data 
is obtained with a~Breit\nobreakdash--Wigner 
line shape when 
the~detector resolution is included. 
Mass parameters for the~$\chicone(3872)$
and $\psitwos$~signals are allowed
to vary in the fit, 
while the~mass difference is
    Gaussian
constrained to 
the~known value~\cite{LHCb-PAPER-2020-009}.
The~width parameters of 
the~$\chicone(3872)$ and $\psitwos$~states 
are fixed to known 
values~\cite{
LHCb-PAPER-2020-009,PDG2022}
using a~Gaussian constraint. 
The~detector resolution functions are described by 
a~symmetric modified Gaussian function 
with power\nobreakdash-law tails 
on both sides of the~distribution~\cite{LHCb-PAPER-2011-013,
Skwarnicki:1986xj}, 
with all parameters determined from simulation.
The~resolutions are further corrected 
by a~common scale factor, $s_{\jpsi\pip\pim}$, 
that accounts for a~small discrepancy 
between data and simulation
and is allowed to vary in the~fit.

\begin{table}[t]
	\centering
	\caption{
        Signal yields, hadron masses,   
        and detector resolution scale 
	factors 
	from the~simultaneous 
        fit described in the~text.
        The~parameters $m_{\Bs}$,
        $s_{\Bs}$  and $s_{\jpsi\pip\pim}$
        are shared in the fit 
        among the~two mass ranges.
	The~uncertainties are statistical only.}
	\label{tab:sim_res}
	\vspace{2mm}
	\begin{tabular*}{0.80\textwidth}{@{\hspace{3mm}}l@{\extracolsep{\fill}}lcc@{\hspace{2mm}}}
      \multicolumn{2}{l}{Parameter} 
     & $\decay{\Bs}{\chicone(3872)\pip\pim}$
     & $\decay{\Bs}{\psitwos\pip\pim}$ 
   \\[1.5mm]
  \hline 
  \\[-1.5mm]
   $N$  
   & 
   & $\phantom{00}155  \pm 23$
   & $\phantom{0}1301 \pm 47$  
   \\
   $m_{\chicone(3872)}$
   & $\left[\!\mevcc\right]$
   & $3871.57 \pm 0.09$
   & --- 
   \\
   $m_{\psitwos}$
   & $\left[\!\mevcc\right]$
   & --- 
   & $3686.08 \pm 0.07$
   \\
   $m_{\Bs}$ & $\left[\!\mevcc\right]$ &  
  \multicolumn{2}{c}{$5366.97 \pm 0.23$}
  \\
  $s_{\Bs}$ & & \multicolumn{2}{c}{ $\phantom{000}1.06 \pm   0.03$   }  
  \\
  $s_{\jpsi\pip\pim}$ 
  &
  & \multicolumn{2}{c}{ $\phantom{000}1.12 \pm      0.03$ } \\
   \end{tabular*}
\end{table}

The~fit is performed
using the~\Bs~mass and the~resolution scale factors, 
$s_{\Bs}$ and $s_{\jpsi\pip\pim}$,  
as shared parameters
    between the~\mbox{$\chicone(3872)\pip\pim$}
    and \mbox{$\psitwos\pip\pim$}~final states.
The~\mbox{$\jpsi\pip\pip\pim\pim$} and 
\mbox{$\jpsi\pip\pim$}~mass distributions 
together with 
projections of the~fit 
are shown in Fig.~\ref{fig:fit_sim_chi}.
The~parameters of interest, 
namely the~\Bs~yields, 
masses of the~\Bs~meson,
$\chicone(3872)$~and  $\psitwos$~states,  
and the~resolution scale factors 
are listed in Table~\ref{tab:sim_res}.
The~statistical significance 
for the~\mbox{$\decay{\Bs}
{\chicone(3872)\pip\pim}$}~signal is 
estimated 
using Wilks' theorem~\cite{Wilks:1938dza}
to be  
7.3~standard deviations. 

The ratio of branching fractions $\mathcal{R}$, 
defined in Eq.~\eqref{eq:r} 
is calculated as 
\begin{equation}\label{eq:rcalcone}
\mathcal{R}  
 = 
 \dfrac { N_{\decay{\Bs}{\chicone(3872) \pip\pim}}}
        { N_{\decay{\Bs}{\psitwos \pip\pim}}}
  \times 
  \dfrac { \varepsilon_{\decay{\Bs}{\psitwos \pip\pim}}}
         { \varepsilon_{\decay{\Bs}{\chicone(3872) \pip\pim}}}\,,
\end{equation}
where the~signal yields, 
$N_{\decay{\Bs}{\chicone(3872)\pip\pim}}$
and $N_{\decay{\Bs}{\psitwos\pip\pim}}$,
are taken from 
Table~\ref{tab:sim_res}
and 
$\varepsilon_{\decay{\Bs}{\chicone(3872)\pip\pim}}$ and 
$\varepsilon_{\decay{\Bs}{\psitwos\pip\pim}}$ are 
the~efficiencies 
for the~\mbox{$\decay{\Bs}{\chicone(3872)\pip\pim}$} and 
\mbox{$\decay{\Bs}{\psitwos\pip\pim}$}~decays, respectively. 
The~efficiencies are defined as the~product of 
the~detector geometric acceptance and the  
reconstruction, selection,  
particle 
identification and trigger efficiencies.
All~of the~efficiency contributions, 
except the~particle identification efficiency, 
are determined using simulated samples.
The~efficiencies of the~hadron identification 
are obtained as a~function of particle momentum,
pseudorapidity and number of charged
tracks in the~event using dedicated 
calibration samples of 
\mbox{$\decay{\Dstarp}{ 
\left(\decay{\Dz}{\Km\pip}\right)\pip}$} 
and \mbox{$\decay{\KS}{\pip\pim}$}~decays 
selected in data~\cite{LHCb-DP-2012-003,
LHCb-DP-2018-001}.
 The~efficiency ratio is found to be
 \begin{equation*}\label{eq:effratio}
     \dfrac { \varepsilon_{\decay{\Bs}{\psitwos \pip\pim}}}
         { \varepsilon_{\decay{\Bs}{\chicone(3872) \pip\pim}}} = 
         0.57 \pm 0.01 \,, 
 \end{equation*}
 where the~uncertainty is due to the~limited 
 size of the~simulated samples. 
 The~efficiency ratio differs from unity 
 due to 
 the~different \pt~spectra of pions 
 in 
 the~\mbox{$\decay{\psitwos}
 {\jpsi\pip\pim}$}
 and 
 the~\mbox{$\decay{\chicone(3872)}
 {\jpsi\pip\pim}$}~decays.
 The~resulting value of $\mathcal{R}$~is
 \begin{equation}
     \mathcal{R} = \left( 6.8 \pm 1.1\right)\times 10^{-2}\,,
 \end{equation}
where the~uncertainty is statistical.

\section{Dipion mass spectrum}
\label{sec:peak}

\begin{figure}[t]
  \setlength{\unitlength}{1mm}
  \centering
  \begin{picture}(160,60)
    %
    \put(  0, 0){ 
      \includegraphics*[width=80mm,
      ]{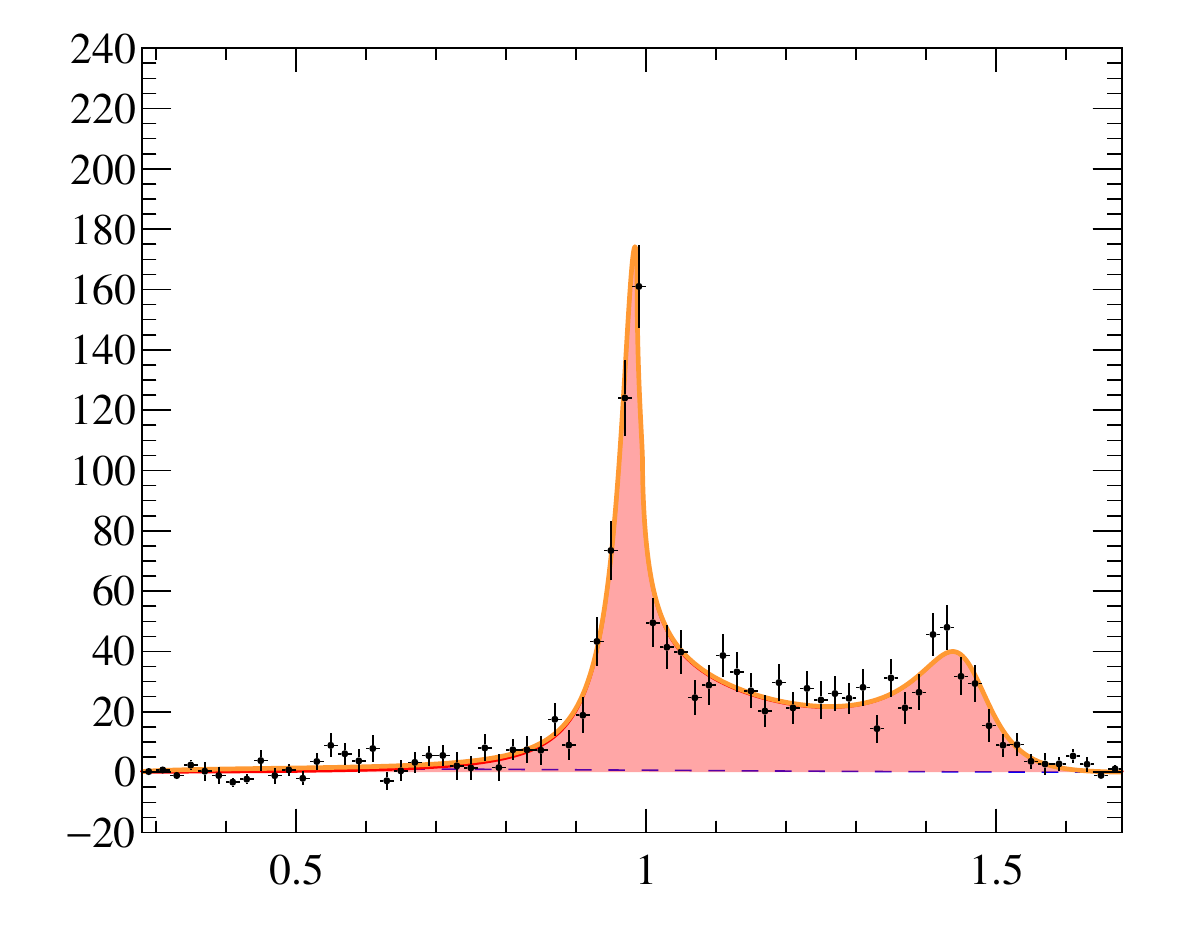}
    }
     \put( 80, 00){ 
      \includegraphics*[width=80mm,
       ]{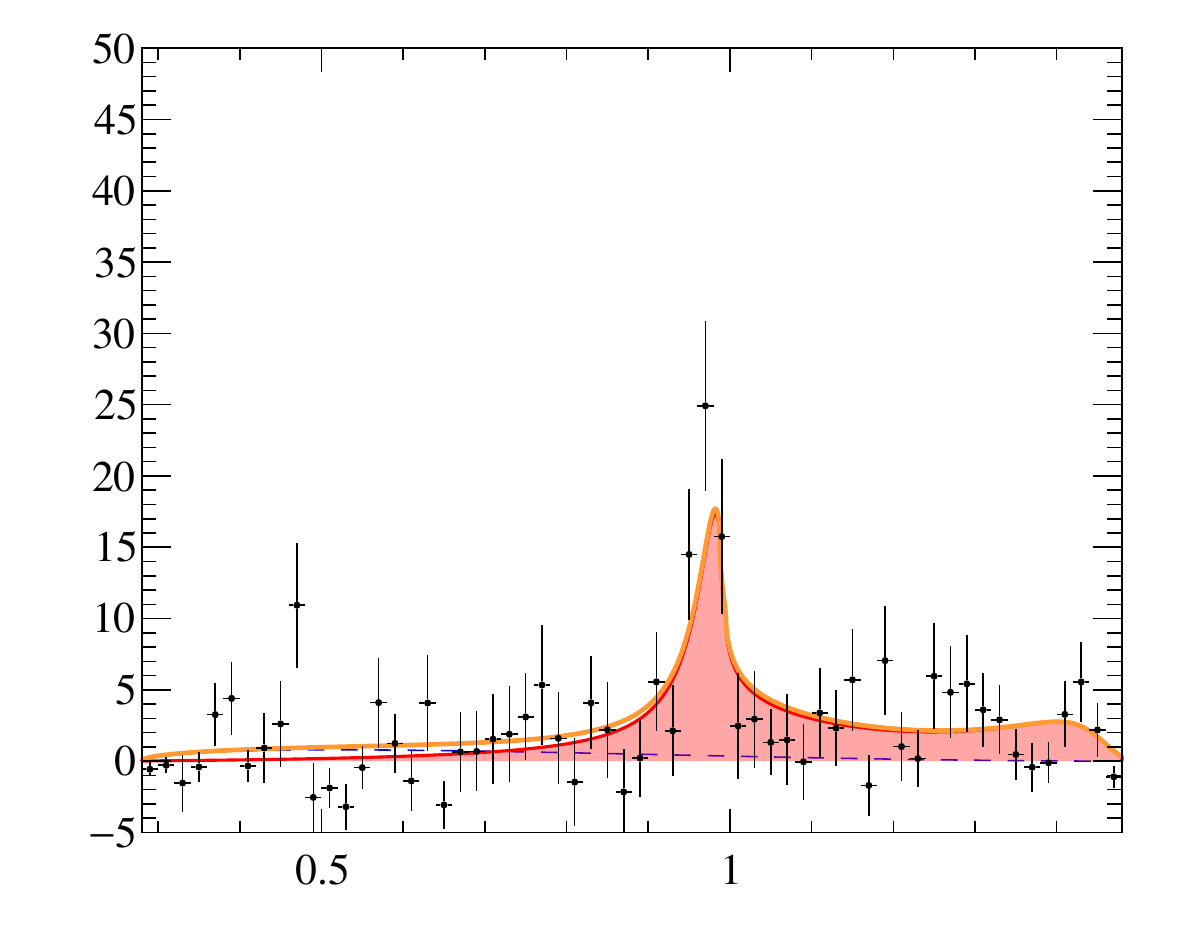}
    }
    \put( 81,42) {\rotatebox[]{90}
    {Yield\,$/\,(20\mevcc)$} }
    \put(  0,42) {\rotatebox[]{90}
    {Yield\,$/\,(20\mevcc)$} }
    \put(125, 0){$m_{\pip\pim}$} 
    \put(143, 0){$\left[\!\gevcc\right]$} 
    \put( 40, 0){$m_{\pip\pim}$} 
    \put( 63, 0){$\left[\!\gevcc\right]$} 
    \put( 60, 49){$\begin{array}{l}\lhcb\\ 9\invfb\end{array}$}
    \put(140, 49){$\begin{array}{l}\lhcb\\ 9\invfb\end{array}$}
    \put(14,51){\scriptsize$\begin{array}{cl} 
    \!\bigplus\mkern-5mu & \text{data} 
    \\
    \begin{tikzpicture}[x=1mm,y=1mm]\filldraw[fill=red!35!white,draw=red,thick]  (0,0) rectangle (5,3);\end{tikzpicture} 
    & \decay{\Bs}{\psitwos \Pf_{0}}
    \\ 
    \color[RGB]{85,83,246} {\hdashrule[0.0ex][x]{5mm}{1.0pt}{2.0mm 0.3mm} } 
    & \decay{\Bs}{\psitwos \pip \pim }
    \\
    \color[RGB]{255,153,51} {\rule{5mm}{2.0pt}}
    & \text{total} 
    \end{array}$
    }   
    \put(94,48){\scriptsize$\begin{array}{cl} 
    \!\bigplus\mkern-5mu & \text{data} 
    \\
    \begin{tikzpicture}[x=1mm,y=1mm]\filldraw[fill=red!35!white,draw=red,thick]  (0,0) rectangle (5,3);\end{tikzpicture} 
    &  \decay{\Bs}{\chicone(3872) \Pf_{0}}
    \\ 
    \color[RGB]{85,83,246} {\hdashrule[0.0ex][x]{5mm}{1.0pt}{2.0mm 0.3mm} } 
    & \decay{\Bs}{\chicone(3872) \pip \pim }
    \\
    \color[RGB]{255,153,51} {\rule{5mm}{2.0pt}}
    & \text{total} 
    \end{array}$
    }
  \end{picture}
	\caption {\small
The~background-subtracted mass spectra for the~dipion system 
recoiling against the~$\chicone(3872)$ or $\psitwos$~states
for (left)~\mbox{$\decay{\Bs}{\psitwos\pip\pim}$}
and (right)~\mbox{$\decay{\Bs}{\chicone(3872)\pip\pim}$}~decays.
The~results of the~fit, described in the~text, are overlaid.}
	\label{fig:xcf_fit}
\end{figure}

The~mass spectra for the~dipion system 
recoiling against the~$\chicone(3872)$ and $\psitwos$~states
are obtained using the~\sPlot technique~\cite{Pivk:2004ty}, 
based on results of the two-dimensional 
fit described in previous section. 
The~spectra shown in Fig.~\ref{fig:xcf_fit}
exhibit significant deviations from 
the~phase\nobreakdash-space distribution 
and are 
similar to 
the~$\pip\pim$~mass spectrum 
observed in 
the~\mbox{$\decay{\Bs}
{\jpsi\pip\pim}$}~decay~\cite{LHCb-PAPER-2011-002,
LHCb-PAPER-2011-031,LHCb-PAPER-2013-069}
and 
the~S\nobreakdash-wave 
$\pip\pim$~component in 
\mbox{$\decay{\Ds}
{\pip\pip\pim}$}~decays~\cite{BESIII:2021jnf,
LHCb-PAPER-2022-030}.

In~Ref.~\cite{LHCb-PAPER-2011-002} it has been found 
that the~$\pip\pim$~mass spectrum from 
\mbox{$\decay{\Bs}{\jpsi\pip\pim}$}~decays 
can be described by a~coherent sum of
the~\mbox{$\Pf_0(980)$}~amplitude 
and a contribution 
from a~high\nobreakdash-mass~scalar meson, 
initially identified as  
the~\mbox{$\Pf_0(1370)$}~state.
References~\cite{Ochs:2013vxa,Ochs:2013gi}
suggest a~better interpretation of 
the~high\nobreakdash-mass state as 
the~\mbox{$\Pf_0(1500)$}~resonance. 
A~dedicated amplitude analysis 
from Ref.~\cite{LHCb-PAPER-2013-069}
followed this suggestion and confirmed 
the~dominant 
\mbox{$\Pf_0(980)$}
and~\mbox{$\Pf_0(1500)$}~contributions
in \mbox{$\decay{\Bs}{\jpsi\pip\pim}$}~decays. 
Assuming a~similarity with 
the~\mbox{$\decay{\Bs}{\jpsi\pip\pim}$}~decays, 
the~mass~spectra of the~dipion system 
recoiling against 
the~$\chicone(3872)$ and $\psitwos$ 
states are described with a~function consisting 
of two components:
\begin{itemize}
    \item A component corresponding to a~coherent 
    sum of the~scalar 
    $\Pf_0(980)$ and $\Pf_0(1500)$~amplitudes 
    and parameterised as 
    \begin{equation} \label{eq:fpipi}
   F( m  )  \propto 
  m
  q p^3 
 \left| f\mathcal{A}_{\Pf_0(980)} ( m ) + 
 \mathrm{e}^{i\varphi} \mathcal{A}_{\Pf_0(1500)}( m ) 
 \right|^2 \,,
    \end{equation}
where $m$ 
is 
the~$\pip\pim$~mass, 
$q$~is the~momentum of the~$\pip$~meson
in the~$\pip\pim$~rest frame, 
$p$~is the~momentum of 
the~$\pip\pim$ system 
in the~$\Bs$~rest frame, 
$\mathcal{A}_{\Pf_0(980)}$ 
and $\mathcal{A}_{\Pf_0(1500)}$ 
are the~$\Pf_0(980)$ 
and $\Pf_0(1500)$~amplitudes,
$\varphi$ is a~relative phase 
and the~real coefficient $f$~characterises 
the~relative contributions of 
the~$\Pf_0(980)$ 
and $\Pf_0(1500)$~components. 
The~amplitude $\mathcal{A}_{\Pf_0(1500)}$  
is parameterised as 
a~relativistic S\nobreakdash-wave 
Breit\nobreakdash--Wigner function, 
while 
the~modified 
Flatt\'e\nobreakdash--Bugg 
amplitude~\cite{Flatte:1976xu,Bugg:2008ig}
\,(see Eq.~(18) in Ref.~\cite{LHCb-PAPER-2013-069})
is used for the~$\Pf_0(980)$~state.
\item A~component corresponding to 
incoherent 
nonresonant 
contribution 
and parameterised by 
the~$\Phi_{2,3}(m)$~phase\nobreakdash-space 
function~\cite{Byckling}. 
\end{itemize}
The~fit is performed simultaneously 
for the~dipion mass 
spectra from 
the~\mbox{$\decay{\Bs}{\chicone(3872)\pip\pip}$}
and \mbox{$\decay{\Bs}{\psitwos\pip\pip}$}~decays. 
The~shape parameters of the~$\Pf_0(980)$ and $\Pf_0(1500)$
states are shared in the~fit.
To~stabilise the~fit, 
Gaussian 
constraints 
are applied
to 
the~parameters of the~$\Pf_0(980)$~state 
and 
the~mass and width 
of the~$\Pf_0(1500)$~state, 
according to 
Solution\,I from Ref.~\cite{LHCb-PAPER-2013-069}.
The~fit results are
shown 
in Fig.~\ref{fig:xcf_fit}.
This~simplistic model 
qualitatively describes 
the~major contributions to the~dipion mass spectrum 
from 
\mbox{$\decay{\Bs}{\psitwos\pip\pim}$}~decays
and supports the~hypothesis of the~dominant contribution
of two S\nobreakdash-wave resonances.  
The~fit indicates the~necessity of 
a~dedicated analysis 
to properly account for the~sub\nobreakdash-leading contributions.
The~same model, consisting of  
two coherent contributions
from the~$\Pf_0(980)$ and $\Pf_0(1500)$~states 
describes well the~dipion mass spectrum 
from the~\mbox{$\decay{\Bs}
{\chicone(3872)\pip\pim}$}~decay.
The~statistical significance for 
the~ \mbox{$\decay{\Bs}
{\chicone(3872)\Pf_0(980)}$}~decay 
is 
estimated using Wilks' theorem and found 
to be~$9.1$~standard 
deviations.

\section{Systematic uncertainties}
\label{sec:sys}

Due to the~similar decay topologies, 
systematic uncertainties largely cancel 
in the~ratio~$\mathcal{R}$. 
The~remaining contributions to systematic uncertainties
are summarized in Table~\ref{tab:sys_ratios}  
and discussed below.

\begin{table}[b]
	\centering
	\caption{Relative systematic uncertainties (in \%) for 
the~ratio of branching fractions. 
The~sources are described in the~text. } 
		\label{tab:sys_ratios}
	\vspace*{2mm}
    \begin{tabular*}{0.50\textwidth}{@{\hspace{3mm}}l@{\extracolsep{\fill}}c@{\hspace{2mm}}}
    Source  
    & $\upsigma_{\mathcal{R}}\,\left[\%\right]$
     \\[1mm]
    \hline 
    \\[-2.5mm]
  Fit model   
  & 2.5
  \\
  $\Bs$ decay model
  & 0.9

  \\
  Efficiency corrections
  & 0.1
  \\
  Trigger efficiency
  &  1.1 
  \\
  Data\nobreakdash-simulation difference
  & 2.0 
  \\
  Simulated sample size 
  & 0.5

  \\[1mm] 
  \hline 
  \\[-2.5mm]
    Sum in quadrature 
    & 3.5 
\end{tabular*}	
	\vspace*{3mm}
\end{table}

The systematic uncertainties arising from 
imperfect knowledge
of the~signal and background shapes 
used for determination of 
the~\mbox{$\decay{\Bs}{\chicone(3872)\pip\pim}$}
and \mbox{$\decay{\Bs}{\psitwos\pip\pim}$}~signal yields 
are estimated with alternative models.
For the~\Bs~signal shape the~bifurcated 
Student's $t$\nobreakdash-distribution~\cite{Student}, 
Apollonios  and Hypatia distributions~\cite{Santos:2013gra}
are tested as alternative models. 
The~$\jpsi\pip\pim$~mass resolution functions
are also modelled with  Student's $t$\nobreakdash-distribution 
and symmetric Apollonios functions.
The~degree of the polynomial functions 
used for the~parameterisation of 
the~background and 
the~non\nobreakdash-resonant 
\mbox{$\jpsi\pip\pim$}~components
is varied by one unit. 
Exponential functions are used as components in 
alternative background models.  
In addition,  
the~detector resolution 
scale factors $s_{\Bs}$
and $s_{\jpsi\pip\pim}$
are constrained to 
values of 
\mbox{$s_{\Bs}=1.04\pm0.02$},
\mbox{$s_{\jpsi\pip\pim}=1.06\pm0.02$}
from Ref.~\cite{LHCb-PAPER-2020-035}
and 
\mbox{$s_{\Bs}=1.052\pm0.003$}, 
\mbox{$s_{\jpsi\pip\pim}=1.048\pm0.004$}
from Ref.~\cite{LHCb-PAPER-2020-009}
using Gaussian constraints.
For each alternative model 
the~ratio of
the~\mbox{$\decay{\Bs}{\chicone(3872)\pip\pim}$}
and \mbox{$\decay{\Bs}{\psitwos\pip\pim}$}~signal yields 
is recalculated. 
The~maximum relative deviation 
with respect to 
the~baseline value is  found to  be 2.5\%  
which is assigned as a~relative
systematic uncertainty for 
the~ratio~$\mathcal{R}$. 
The~fit procedure itself 
is tested using a~large sample of  
pseudoexperiments, generated using the~default model
with parameters extracted from  data and the~results 
are found to be unbiased.

The~\mbox{$\decay{\Bs}{\chicone(3872)\pip\pim}$}
and \mbox{$\decay{\Bs}{\psitwos\pip\pim}$}~decays 
are simulated as phase\nobreakdash-space decays and corrected 
to reproduce the~dipion mass spectra
observed in data.   
A~weighting procedure, based 
on a gradient boosted tree 
algorithm~\cite{Rogozhnikov:2016bdp},
is used for corrections of simulated samples. 
The systematic uncertainty related to the~correction 
method is estimated by varying 
the~hyper\nobreakdash-parameters  
of the~regression trees ensemble. 
The~maximum deviation from the baseline value, 0.9\%, 
is taken as the uncertainty
associated with the~unknown \Bs~decay models. 

An~additional systematic uncertainty on 
the~ratio~$\mathcal{R}$
arises due to
differences between data and  simulation. 
In particular, there are differences in 
the~reconstruction efficiency 
of charged\nobreakdash-particle tracks that 
do not cancel completely in the~ratio due 
to the~different kinematic distributions 
of the~final\nobreakdash-state particles.
The~track\nobreakdash-finding efficiencies 
obtained from 
simulated samples 
are corrected \mbox{using}
data\nobreakdash-driven techniques~\cite{LHCb-DP-2013-002}.
The~uncertainties related to~the~efficiency 
correction factors, 
\mbox{together}
with the~uncertainty on the~hadron\nobreakdash-identification
efficiency due to the~finite size of 
the~calibration 
samples~\mbox{\cite{LHCb-DP-2012-003, LHCb-DP-2018-001}},
are propagated to
    the~ratio
of 
the~total efficiencies using pseudoexperiments
and amount to~0.1\%.

A~systematic uncertainty on
the~ratio
related to 
the knowledge of the~trigger efficiencies is estimated 
by comparing the~ratios of trigger
efficiencies in data and simulation for large samples 
of~$\decay{\Bp}{\jpsi\Kp}$  and 
$\decay{\Bp}{\psitwos\Kp}$~decays~\cite{LHCb-PAPER-2012-010}
and is taken to be 1.1\%.

Remaining  data\nobreakdash-simulation differences, 
that are not previously discussed, 
are investigated by varying 
the~selection criteria.
The~resulting variations in 
the~ratios of 
the~efficiency\nobreakdash-corrected yields  
do not exceed 2\%, which  
is taken as the corresponding systematic 
uncertainty.
The~final systematic uncertainty 
considered on the~ratios of
branching fractions is 
due to the~knowledge of  
the~ratios of efficiencies in 
Eq.~\eqref{eq:rcalcone}, 
limited by the~size of simulated samples.
It~is determined to be 0.5\%.

No systematic uncertainty
is included for the~admixture
of the~\CP-odd and \CP-even 
\Bs~eigenstates in 
the~\mbox{$\decay{\Bs}{\chicone(3872)\pip\pim}$}
and \mbox{$\decay{\Bs}
{\psitwos\pip\pim}$}~decays~\cite{DeBruyn:2012wj}, 
which is assumed to be the~same for both channels. 
Analysis of the~dipion spectrum from the~\mbox{$\decay{\Bs}
{\psitwos\pip\pim}$}~decays in Sec.~\ref{sec:peak}
indicates 
that the~final 
state is predominantly \CP-odd, 
with the~effective lifetime 
corresponding to 
the~heavy\nobreakdash-mass 
long\nobreakdash-lifetime 
\Bs~eigenstate~\cite{LHCb-PAPER-2012-017,
D0:2016nbv,
CMS:2017ygm}.
The~dipion spectrum from the~\mbox{$\decay{\Bs}
{\chicone(3872)\pip\pim}$}~decays
is also consistent with being predominantly \CP-odd.
In~the~extreme case that 
the~\mbox{$\decay{\Bs}
{\chicone(3872)\pip\pim}$}~decay
is an~equal 
mixture of the~short\nobreakdash-lifetime and 
long\nobreakdash-lifetime eigenstates, 
the~corresponding ratio 
of branching fractions 
would change by~2.4\,\%.

The~statistical significance 
for the~\mbox{$\decay{\Bs}{\chicone(3872)\pip\pim}$}~decay
is recalculated using Wilks' theorem 
for each alternative fit model, 
and the~smallest value of 
$7.3$~standard deviations
is taken as the~significance 
that includes 
the~systematic uncertainty.

Alternative models are used also for 
the~fit to the~dipion mass spectra 
from the~\mbox{$\decay{\Bs}{\chicone(3872)\pip\pim}$}
and \mbox{$\decay{\Bs}{\psitwos\pip\pim}$}~channels.
In~particular,
Solution\,II from Ref.~\cite{LHCb-PAPER-2013-069}
is investigated as 
an~alternative external constraint for 
the~$\Pf_0(980)$~and $\Pf_0(1500)$~parameters. 
The~model with 
the~$\Pf_0(1370)$~resonance instead of 
the~$\Pf_0(1500)$~state is used 
as alternative model
with a~constraint to the~known parameters 
of the~$\Pf_0(1370)$~state~\cite{PDG2022}.
As an~alternative model for 
the~noncoherent 
nonresonant component, 
a~product of 
the~$\Phi_{2,3}$~phase\nobreakdash-space function and 
the~positive 
second\nobreakdash-order polynomial 
function~\cite{karlin1953geometry}
has been probed.
Also, a~coherent nonresonant contribution,
parameterised with a~complex\nobreakdash-valued 
constant function, 
is added to the~function from 
Eq.\,\eqref{eq:fpipi}.
The~smallest significance for 
the~\mbox{$\decay{\Bs}{\chicone(3872)\Pf_0(980)}$}~decay
is found to be $7.4$~standard deviations, which
is taken as the~significance including 
systematic uncertainty.  

\section{Summary}
\label{sec:results}

An~observation of 
the~\mbox{$\decay{\Bs}{\chicone(3872)\pip\pim}$}~decays 
with significance exceeding 
7~standard deviations 
is reported 
using 
proton\nobreakdash-proton
collision  data, 
corresponding to 
integrated luminosities of 1, 2 and 6\invfb, 
collected by the~LHCb experiment
at centre\nobreakdash-of\nobreakdash-mass
energies of 7, 8 and 13\tev, respectively. 
Using the~\mbox{$\decay{\Bs}{\psitwos\pip\pim}$}~decay 
as a~normalization channel,
and neglecting 
the~possible interference effects, 
the~ratio of branching fractions
is measured to be 
\begin{equation*}
\dfrac{ \BR\left(\decay{\Bs}
{ \chicone(3872)  \pip\pim}\right) 
\times \BR\left(\decay{\chicone(3872)}{\jpsi\pip\pim}\right)}
{\BR\left(\decay{\Bs}
{ \psitwos  \pip\pim}\right) 
\times \BR\left(\decay{\psitwos}{\jpsi\pip\pim}\right)}
 = \left( 6.8  \pm 1.1 \pm 0.2 \right) \times 10^{-2}\,,
\end{equation*}
where the~first uncertainty is statistical 
and the~second systematic.
The~measured ratio of the~branching fractions 
exceeds by a~factor of three 
the~analogous ratio for 
the~\mbox{$\decay{\Bs}{\chicone(3872)\Pphi}$} 
and \mbox{$\decay{\Bs}{\psitwos\Pphi}$}~decays 
reported in Refs.~\cite{Sirunyan:2020qir,
LHCb-PAPER-2020-035}.
Using the~known branching
fractions 
for the~\mbox{$\decay{\Bs}
{\psitwos\pip\pim}$},
and \mbox{$\decay{\psitwos}
{\jpsi\pip\pim}$}~decays~\cite{PDG2022},
the~branching fraction for
the~\mbox{$\decay{\Bs}
{\left( \decay{\chicone(3872)}{\jpsi\pip\pim}
\right)\pip\pim}$}~is calculated to be 
\begin{equation*}
\BR\left(\decay{\Bs}
{ \chicone(3872)  \pip\pim}\right) 
\times \BR\left(\decay{\chicone(3872)}
{\jpsi\pip\pim}\right)
=  \left( 1.6 \pm 0.3 \pm 0.1 \pm 0.3 \right) 
\times 10^{-6}\,,
\end{equation*}
where 
the~third uncertainty is
due to imprecise knowledge of 
the branching fractions for 
the~\mbox{$\decay{\Bs}{\psitwos\pip\pim}$}
and \mbox{$\decay{\psitwos}{\jpsi\pip\pim}$}~decays~\cite{PDG2022}.
    The~obtained value is more than twice
    smaller
    than the~branching fractions
    of 
    the~\mbox{$\decay{\Bs}{\left(\decay{\chicone(3872)}{\jpsi\pip\pim}\right)\Pphi}$}
    and
    the~\mbox{$\decay{\Bs}{\left(\decay{\chicone(3872)}{\jpsi\pip\pim}\right)
        \left( \Kp\Km\right)_{\mathrm{non}-\Pphi}}$}~decays~\cite{Sirunyan:2020qir,
      LHCb-PAPER-2020-035,PDG2022}.

The~mass spectrum  of the~dipion system recoiling
against the~$\chicone(3872)$~state 
shows a~similarity with those from 
the~\mbox{$\decay{\Bs}{\psitwos\pip\pim}$}
and \mbox{$\decay{\Bs}
{\jpsi\pip\pim}$}~decays~\cite{LHCb-PAPER-2013-069}, 
compatible  with a~dominant 
S\nobreakdash-wave contribution, 
and exhibits  a~large \mbox{$\decay{\Bs}
{\chicone(3872)\Pf_0(980)}$}~component
with significance 
exceeding 7~standard deviations. 
With a~larger data sample and better understanding of
the~S\nobreakdash-wave 
$\pip\pim$~scattering~\cite{Ropertz:2018stk}, 
a~precise determination of the~$\Pf_0(980)$~component 
in this decay will be possible.

\section*{Acknowledgements}
%
%
\noindent We express our gratitude to our colleagues in the~CERN
accelerator departments for the excellent performance of the~LHC.
We~thank the~technical and administrative staff at the~LHCb
institutes.
We acknowledge support from CERN and from the national agencies:
CAPES, CNPq, FAPERJ and FINEP\,(Brazil); 
MOST and NSFC\,(China); 
CNRS/IN2P3\,(France); 
BMBF, DFG and MPG\,(Germany); 
INFN\,(Italy); 
NWO\,(Netherlands); 
MNiSW and NCN\,(Poland); 
MEN/IFA\,(Romania); 
MICINN\,(Spain); 
SNSF and SER\,(Switzerland); 
NASU\,(Ukraine); 
STFC\,(United Kingdom); 
DOE NP and NSF\,(USA).
We~acknowledge the~computing resources that are provided by CERN,
IN2P3\,(France),
KIT and DESY\,(Germany), INFN\,(Italy), SURF\,(Netherlands),
PIC (Spain), GridPP\,(United Kingdom), 
CSCS\,(Switzerland), IFIN-HH\,(Romania), CBPF\,(Brazil),
Polish WLCG  (Poland) and NERSC\,(USA).
We~are indebted to the~communities behind the~multiple open-source
software packages on which we depend.
Individual groups or members have received support from
ARC and ARDC\,(Australia);
Minciencias\,(Colombia);
AvH Foundation\,(Germany);
EPLANET, Marie Sk\l{}odowska-Curie Actions and ERC\,(European Union);
A*MIDEX, ANR, IPhU and Labex P2IO, and R\'{e}gion Auvergne-Rh\^{o}ne-Alpes\,(France);
Key Research Program of Frontier Sciences of CAS, CAS PIFI, CAS CCEPP, 
Fundamental Research Funds for the~Central Universities, 
and Sci. \& Tech. Program of Guangzhou\,(China);
GVA, XuntaGal, GENCAT and Prog.~Atracci\'on Talento, CM\,(Spain);
SRC\,(Sweden);
the~Leverhulme Trust, the~Royal Society
 and UKRI\,(United Kingdom).



\clearpage 
\addcontentsline{toc}{section}{References}
\bibliographystyle{LHCb}
\bibliography{main,standard,LHCb-PAPER,LHCb-CONF,LHCb-DP,LHCb-TDR}

\newpage
\centerline
{\large\bf LHCb collaboration}
\begin
{flushleft}
\small
R.~Aaij$^{32}$\lhcborcid{0000-0003-0533-1952},
A.S.W.~Abdelmotteleb$^{50}$\lhcborcid{0000-0001-7905-0542},
C.~Abellan~Beteta$^{44}$,
F.~Abudin{\'e}n$^{50}$\lhcborcid{0000-0002-6737-3528},
T.~Ackernley$^{54}$\lhcborcid{0000-0002-5951-3498},
B.~Adeva$^{40}$\lhcborcid{0000-0001-9756-3712},
M.~Adinolfi$^{48}$\lhcborcid{0000-0002-1326-1264},
P.~Adlarson$^{77}$\lhcborcid{0000-0001-6280-3851},
H.~Afsharnia$^{9}$,
C.~Agapopoulou$^{13}$\lhcborcid{0000-0002-2368-0147},
C.A.~Aidala$^{78}$\lhcborcid{0000-0001-9540-4988}
% ...

{\footnotesize \it

% ...
$^{9}$Universit{\'e} Clermont Auvergne, CNRS/IN2P3, LPC, Clermont-Ferrand, France\\
% ...
$^{13}$LPNHE, Sorbonne Universit{\'e}, Paris Diderot Sorbonne Paris Cit{\'e}, CNRS/IN2P3, Paris, France\\
% ...
$^{32}$Nikhef National Institute for Subatomic Physics, Amsterdam, Netherlands\\
% ...
$^{40}$Instituto Galego de F{\'\i}sica de Altas Enerx{\'\i}as (IGFAE), Universidade de Santiago de Compostela, Santiago de Compostela, Spain\\
% ...
$^{44}$Physik-Institut, Universit{\"a}t Z{\"u}rich, Z{\"u}rich, Switzerland\\
% ...
$^{48}$H.H. Wills Physics Laboratory, University of Bristol, Bristol, United Kingdom\\
% ...
$^{50}$Department of Physics, University of Warwick, Coventry, United Kingdom\\
% ...
$^{53}$School of Physics and Astronomy, University of Glasgow, Glasgow, United Kingdom\\
$^{54}$Oliver Lodge Laboratory, University of Liverpool, Liverpool, United Kingdom\\
% ...
$^{62}$Syracuse University, Syracuse, NY, United States\\
% ...
$^{77}$Department of Physics and Astronomy, Uppsala University, Uppsala, Sweden, associated to $^{53}$\\
$^{78}$University of Michigan, Ann Arbor, MI, United States, associated to $^{62}$\\
% ...
}
\end{flushleft}

\end{document}